\documentclass[nonacm,acmownedfalsemanuscript,screen]{acmart}

\AtBeginDocument{%
  \providecommand\BibTeX{{%
    \normalfont B\kern-0.5em{\scshape i\kern-0.25em b}\kern-0.8em\TeX}}}

\setcopyright{acmcopyright}
\copyrightyear{2023}
\acmYear{2023}
\acmDOI{XXXXXXX.XXXXXXX}

\usepackage{amsmath}
\usepackage{amsfonts}
\usepackage{xcolor}
\usepackage{soul}
\usepackage{pdfpages}
\usepackage{fancyvrb}
\usepackage{caption}
\usepackage{subcaption}

\usepackage{graphicx}
\usepackage{float}
\usepackage{makecell}
\usepackage{grffile}
\usepackage[most]{tcolorbox}

\usepackage{amsthm}
\usepackage{enumitem}
\usepackage{scalerel,stackengine}
\usepackage{xspace}
\usepackage{hyperref}

\usepackage{amsopn}
\usepackage{multirow}
\usepackage{tabularx}
\usepackage{hhline}
\usepackage{arydshln}
\usepackage[normalem]{ulem}
\usepackage{tikz}
\usepackage{pgf}
\usepackage{collcell}
\usepackage[utf8]{inputenc}
\usepackage{geometry}

\usepackage{graphbox}
\usepackage{setspace}

\newcolumntype{E}{>{\collectcell\usermacro}c<{\endcollectcell}}
\newcommand\usermacro[1]{\pgfmathparse{10000*#1}\pgfmathprintnumber\pgfmathresult}

\usepackage{algorithmicx}
\usepackage{algorithm}
\usepackage{algpseudocode}

\makeatletter
\patchcmd{\ALG@step}{\addtocounter{ALG@line}{1}}{\refstepcounter{ALG@line}}{}{}
\newcommand{\ALG@lineautorefname}{Step}
\makeatother

\usepackage{listings}

\DeclareMathAlphabet{\mathup}{OT1}{\familydefault}{m}{n}

\DeclareSymbolFont{yhlargesymbols}{OMX}{yhex}{m}{n}
\DeclareMathAccent{\wideparen}{\mathord}{yhlargesymbols}{"F3}
\xdefinecolor{green}{rgb}{0.04, 0.85, 0.32}
\xdefinecolor{darkgreen}{rgb}{0.24, 0.7, 0.44}
\xdefinecolor{mint}{rgb}{0.24, 0.71, 0.54}
\xdefinecolor{officegreen}{rgb}{0.0, 0.5, 0.0}
\xdefinecolor{napiergreen}{rgb}{0.16, 0.5, 0.0}
\xdefinecolor{maroon}{rgb}{0.65,0.06,0.17}
\xdefinecolor{blue}{rgb}{0,0.2,0.6}
\xdefinecolor{phthalogreen}{rgb}{0.07,0.21,0.14}
\definecolor{grassgreen}{RGB}{92,135,39}
\newcommand{\logdet}[1]{\log\det\left(#1\right)}                    
\newcommand{\mat}[1]{\mathbf{{#1}}}                                 
\renewcommand{\vec}[1]{\mathbf{{#1}}}                                 

\DeclareMathOperator*{\argmin}{arg\,min}  
\DeclareMathOperator*{\argmax}{arg\,max}  

\newcommand{\regpenalty}{\alpha}                    
\newcommand{\penaltyfunc}{\Phi}                    
\newcommand{\penaltyfunction}[1]{\penaltyfunc({#1})}                    
\newcommand*{\tran}{^{\mkern-1.5mu\mathsf{T}}}                
\newcommand*{\adj}{^{\mkern-1.5mu\mathsf{*}}}                 
\newcommand*{\inv}{^{\mkern-1.5mu\mathsf{-1}}}                
\newcommand{\pseudoinverse}[1]{\Bigl(#1\Bigr)^{\dagger}}          
\newcommand{\domain}{\mathcal{D}}                             
\newcommand{\trace}{\mathrm{Tr}}                              
\newcommand{\Trace}[1]{\trace \left(#1\right)}                
\newcommand{\wnorm}[2]{\left\| {#1} \right\|_{#2}}            
\newcommand{\sqwnorm}[2]{\left\| {#1} \right\|^2_{#2}}        

\newcommand\restr[2]{{ \left.\kern-\nulldelimiterspace        
                     {#1}\vphantom{\big|} \right|_{#2}}}

\newcommand{\Rnum}{\mathbb{R}}  
\newcommand{\xcont}{u}                             

\newcommand{\x}{\vec{u}}                           

\newcommand{\modelstate}{\x}

\newcommand{\ObsOper}{\mat{O}}

\newcommand{\y}{\mathbf{y}}                        
\newcommand{\obs}{\y}                              

\newcommand{\param}{\vec{\theta}}                  
\newcommand{\iparam}{\param}                       
\newcommand{\iparprior}{\iparam_{\rm pr}}          
\newcommand{\paramprior}{\iparprior}               
\newcommand{\iparb}{\iparprior}                    
\newcommand{\iparpost}{\iparam_{\rm post}^\obs}    
\newcommand{\ipara}{\iparpost}                     

\newcommand{\Nstate}{\textsc{N}_{\rm state}}                    

\newcommand{\Nobs}{\textsc{N}_{\rm obs}}                        
\newcommand{\nobs}{n_{t}}                                  
\newcommand{\nobstimes}{\nobs}                             
\newcommand{\Nsens}{{n_{\rm s}}}                         
\newcommand{\Cparamprior}{\mat\Gamma_{{\rm pr}}}                
\newcommand{\Cparampost}{\mat\Gamma_{{\rm post}}}               
\newcommand{\Cobsnoise}{\mat{\Gamma}_{ {\rm noise}}}            
\newcommand{\Cparampriormat}{\Cparamprior}                      
\newcommand{\Cparampostmat}{\Cparampost}                        

\newcommand{\obsnoise}{\vec{\delta}}                             

\newcommand{\Fcont}{\mathcal{F}}                                 
\newcommand{\F}{\mathbf{F}}                                      

\newcommand{\Sol}{\mathcal{S}}  
\newcommand{\Pred}{\mathcal{P}}                                  
\newcommand{\PredOper}{\Pred}                                  

\newcommand{\obj}{\mathcal{J}}            

\newcommand{\FIM}{\mathsf{FIM}}
\newcommand{\utilityfunc}{\mathcal{U}}

\newcommand{\Prob}{\mathbb{P}}                                   
\newcommand{\CondProb}[2]{\mathbb{P}\left(#1|#2 \right)}  
\newcommand{\GM}[2]{\mathcal{N}\!\left( {#1}, {#2}\right)}       

\newcommand{\like}{\mathcal{L}}                          
\newcommand{\Like}[2]{\like{\left(#1|#2\right)}}                                  

\newcommand{\Expect}[2]{\mathbb{E}_{#1}{\Bigl[ #2 \Bigr]} }     
\newcommand{\baseline}{b}
\newcommand{\hyperparam}{\vec{p}}     
\newcommand{\design}{\boldsymbol{\zeta}}                                       
\newcommand{\binarydesign}{\design^{\rm b}}                                       
\newcommand{\weightfunc}{\omega}                                    
\newcommand{\designmat}{\mat{W}}                                    
\newcommand{\wdesignmat}{\designmat_{\Gamma}}                       

\newcommand{\pred}{\vec{\rho}}                                      

\stackMath
\newcommand\reallywidehat[1]{%
\savestack{\tmpbox}{\stretchto{%
  \scaleto{%
    \scalerel*[\widthof{\ensuremath{#1}}]{\kern-.6pt\bigwedge\kern-.6pt}%
    {\rule[-\textheight/2]{1ex}{\textheight}}
  }{\textheight}%
}{0.5ex}}%
\stackon[1pt]{#1}{\tmpbox}%
}
\newcommand{\dates}{{DATeS}\xspace}
\newcommand{\pyoed}{{PyOED}\xspace}
\newcommand{\code}[1]{\texttt{{#1}}}
\newcommand{\package}[1]{\code{#1}}

\newcommand*{\opt}{^{\mkern-1.5mu\mathsf{opt}}}               
\newcommand*{\true}{^{\mathrm{true}}}             
\newcommand*{\pr}{^{\mathrm{pr}}}             

\newcommand{\commentout}[1]{\iffalse {#1} \fi}
\newcommand{\todo}[1]{\noindent\textcolor{cyan}{TODO: #1\,}}

\newcommand{\ahmed}[1]{\textcolor{officegreen}{Ahmed:
#1}\marginpar{\textcolor{officegreen}{AA}}}

\definecolor{codegreen}{rgb}{0,0.6,0}
\definecolor{codegray}{rgb}{0.5,0.5,0.5}
\definecolor{codepurple}{rgb}{0.58,0,0.82}
\definecolor{backcolour}{rgb}{0.95,0.95,0.92}

\lstdefinestyle{mystyle}{
    backgroundcolor=\color{backcolour},   
    commentstyle=\color{codegreen},
    keywordstyle=\color{magenta},
    numberstyle=\tiny\color{codegray},
    stringstyle=\color{codepurple},
    basicstyle=\ttfamily\footnotesize,
    breakatwhitespace=false,         
    breaklines=true,                 
    captionpos=b,                    
    keepspaces=true,                 
    numbers=left,                    
    numbersep=5pt,                  
    showspaces=false,                
    showstringspaces=false,
    showtabs=false,                  
    tabsize=2
}

\newcommand{\pyoedversion}{v1.0}

\begin{document}

\newcommand{\shorttitle}{\pyoed: DA and OED in Python}
\title{{PyOED}: An Extensible Suite for Data Assimilation and Model-Constrained Optimal Design of Experiments}

\author{Abhijit Chowdhary}
\authornote{
  The first author developed the nonlinear OED pipeline in \pyoed during a major refactor phase and contributed to the numerical experiments and revision of this article.
}

\affiliation{%
  \institution{Mathematics Department, North Carolina State University}
  \city{Raleigh}
  \state{North Carolina}
  \country{USA}
  \postcode{27606}}
\email{achowdh2@ncsu.edu}

\author{Shady E. Ahmed}
\authornote{The second author developed a set of simulation models in \pyoed
    and contributed to the introduction and revision of this article.}


\affiliation{%
  \institution{School of Mechanical and Aerospace Engineering, Oklahoma State University}
  \city{Stillwater}
  \state{Oklahoma}
  \country{USA}
  \postcode{74078}}
\email{shady.ahmed@okstate.edu}

\author{Ahmed Attia}
\authornote{The last/third author is the main developer of \pyoed and  led the development of both \pyoed and this article.}
\affiliation{%
  \institution{Mathematics and Computer Science Division, Argonne National Laboratory}
  \city{Lemont}
  \state{Illinois}
  \country{USA}
  \postcode{60439}}
\email{aattia@anl.gov}

\begin{abstract}
    This paper describes \pyoed, a highly extensible scientific package that enables developing and testing model-constrained optimal experimental design (OED) for inverse problems.
    Specifically, \pyoed aims to be a comprehensive \emph{Python toolkit} for \emph{model-constrained OED}. 
    The package targets scientists and researchers interested in understanding the details of OED formulations and approaches. 
    It is also meant to enable researchers to experiment with standard and innovative OED technologies with a wide range of test problems (e.g., simulation models). 
    OED, inverse problems (e.g., Bayesian inversion), and data assimilation (DA) are closely related research fields, and their formulations overlap significantly. 
    Thus, \pyoed is continuously being expanded with a plethora of Bayesian inversion, DA, and OED methods as well as new scientific simulation models, observation error models, and observation operators. 
    These pieces are added such that they can be permuted to enable testing OED methods in various settings of varying complexities. The \pyoed core is completely written in Python and utilizes the inherent object-oriented capabilities; however, the current version of \pyoed is meant to be extensible rather than scalable.
    Specifically, \pyoed is developed to ``enable rapid development and benchmarking of OED methods with minimal coding effort and to maximize code reutilization.'' 
    This paper provides a brief description of the \pyoed layout and philosophy and  provides a set of exemplary test cases and tutorials to demonstrate the potential of the package.
\end{abstract}

\ccsdesc[500]{Mathematics of computing~Mathematical software}

\keywords{
    Optimal experimental design, 
    OED,
    inverse problems,
    data assimilation
}

\maketitle


\section{Introduction}
\label{sec:Introduction}
  Recently, interest has increased in developing scalable data assimilation (DA) and uncertainty quantification methodologies for solving large-scale inverse problems. An inverse problem refers to the retrieval of a quantity of interest (QoI) associated with or stemming from a physical phenomenon underlying partial noisy experimental or observational data of that physical system~\cite{aster2018parameter,vogel2002computational,stuart2010inverse}.
  The QoI could be, for example, the model state, initial condition, or other physics quantity. Inverse problems are prominent in a wide spectrum of applications including power grids and atmospheric numerical weather prediction~\cite{ghil1991data,smith2013uncertainty}.
  In these problems, the prediction of the physical phenomena is often formulated as an initial value problem, while the initial condition of the simulator is corrected by fusing all available information. 
  Algorithmic approaches for solving inverse problems seek either a single-point estimate of the target QoI or a full probabilistic description of the knowledge about the QoI given all available information.
  In the former approach seeking a single QoI estimate, the solution of an inverse problem is obtained by solving an optimization problem with an objective to minimize the mismatch between observational data and model simulations, possibly regularized by prior knowledge and uncertainty models.
  The latter approach, commonly known as Bayesian inversion, seeks to characterize the probability distribution of the QoI through the posterior formulated by applying Bayes' rule, that is, the probability distribution of the QoI conditioned by all available information. 
  
  DA methods~\cite{bannister2017review,daley1993atmospheric,navon2009data,smith2013uncertainty,ghil1991data,attia2015hmcfilter,attia2015hmcsampling,attia2015hmcsmoother,attia2017reduced,attia2019dates} 
  aim to efficiently solve large- to extreme-scale inverse problems. They work by fusing 
  information obtained from multiple sources, such as the dynamical model, prior knowledge, noisy and incomplete measurements, and error models, in order to better estimate the state and parameters of the physical system.
  This estimate improves the predictability of the simulation systems developed to make future predictions about the physical phenomena of interest.
  
  The quality of DA systems, and hence the accuracy of their predictions, is heavily influenced by the extent to which the mathematical assumptions reflect reality and depends on the quality of the collected measurements. 
  Optimal data acquisition is the problem of determining the optimal observational strategy, for example, from a large set of candidate observational schemes. 
  This problem is widely formulated as an optimal experimental design (OED) 
  problem~\cite{fedorov2000design,pukelsheim2006optimal}, where the design parameterizes and thus determines the observational configuration. 
  In an OED problem, a design is defined to characterize a candidate configuration or a control, and the quality of the design is quantified by using a utility function.
  The optimal design is then defined as the one that maximizes this utility function or, equivalently, minimizes some OED criterion~\cite{attia2022optimal}.
  Since the aim of Bayesian inference is to estimate the QoI posterior, Bayesian OED seeks an observational configuration that, when combined with the underlying dynamics, would maximize information gain from the data or minimize the posterior uncertainty. 
  Thus, an optimal design is found by solving an optimization problem with the objective to maximize a utility function that quantifies the quality of the design and its influence on the solution of the inverse problem. 
  OED for inverse problems has experienced a recent surge in interest by the scientific computing community; see, for example,~\cite{alexanderian2021optimal} and references therein.

  Numerical testing and experiments are critical for developing efficient OED formulations and algorithms. This process is elementary for successful scientific research in general. Although statisticians have developed a plethora of mathematical formulations and algorithmic approaches for general-purpose OED algorithms, most of the available and publicly accessible OED software tools are 
  limited to idealized formulations and specific applications such as finding optimal collocation points for regression problems or designing clinical experiments. 
  While  numerous  OED methodologies have been developed for model-based applications, open-source software for model-based OED applications is lacking~\cite{flassig2018model}.
  Some of these packages are written in MATLAB, which can limit code reutilization and accessibility in some cases; see, for example, ~\cite{foracchia2004poped}.
  The statistical community provides several OED tools written by experts in the  \textrm{R} programming language to enable solving a wide range of specialized OED problems.
  For example, \package{OptimalDesign}~\cite{harman2019brief} is an OED toolbox developed for regression experiments with uncorrelated observations. 
  The \package{OPDOE} package described in~\cite{rasch2011optimal} enables optimizing locations and number of replications for a polynomial regression design. 
  \package{AlgDesign}~\cite{wheeler2019package} enables finding exact or approximate algorithmic designs that optimize one of three optimality criteria, namely, A-, D-, or I-optimality, using Federov’s exchange algorithm~\cite{fedorov2013theory}. 
  The \package{crossdes} package~\cite{sailer2005crossdes} is a specialized package for OED in crossover studies, 
  for example, in medicine when subjects receive a sequence of different treatments in a longitudinal study;
  see e.g.,~\cite{sibbald1998understanding}.
  \package{EFCOSS}~\cite{rasch2009software} provides an interactive interface and automatic differentiation tools for gradient-based optimization, parameter estimation, and model-based OED.
  \package{GPdoemd}~\cite{olofsson2019gpdoemd} utilizes data-driven Gaussian process surrogate models for model discrimination for optimizing science-based models.
  \package{AutoOED}~\cite{tian2021autooed} is a tools for OED that leverages machine learning for finding optimal design in various applications.
  \package{Pydex}~\cite{kusumo2022risk} and \package{PyApprox}~\cite{jakeman2022pyapprox} have been recently released for OED in scientific simulations and inference problems. 
  \package {Pydex} uses a conditional-value-at-risk design criterion to balance the average information content and risk in inference problems.
  \package{PyApprox} provides tools for probabilistic analysis of scientific simulations, constructing surrogates, sensitivity analysis, 
  Bayesian inference, OED, and forward uncertainty quantification. 
  These tools provide a wide range of OED capabilities. They do not, however, interface easily to scientific simulations, DA algorithms, and inverse problems; and they are not easily adaptable for rapid developments of new methodologies.
  We believe that a general-purpose open-source software package for developing new algorithmic approaches for model-constrained OED problems is still lacking. 
  We believe that such capability is needed and indeed essential for an increasingly large scientific community focusing on DA and model-constrained  OED~\cite{attia2018goal,alexanderian2016fast,huan2013simulation,bui2013computational,flath2011fast,haber2008numerical,haber2009numerical,pukelsheim2006optimal,pronzato2013design}. 
  As a first step in alleviating this limitation, we present \pyoed, 
  a highly extensible open-source Python package written mainly to enable computational scientists to formulate and rapidly test OED---as well as DA---formulations and algorithmic approaches.

  \pyoed is unique in several ways. 
  First, to the best of our knowledge, it is the first open-source scientific computing package for the DA and OED stack designed to provide flexible interfaces for the intermediate structures.
  In particular, \pyoed is designed in an object-oriented programming (OOP) fashion, which enables practitioners to reconfigure and reuse the individual building blocks. 
  Moreover, it is easy to combine \pyoed with other user-defined routines, such as numerical integration of simulation models and optimization routines, thus making \pyoed highly extensible and adaptable to a wide range of applications. 
  Second, it is written in Python, which is arguably the most popular and adopted programming language for recent algorithmic developments in the computational science disciplines.
  It has a huge ecosystem, and the learning curve is relatively smoother than that of other lower-level programming languages such as C/C++. 
  Third, \pyoed is not constrained to a specific inverse problem formulation. 
  Its interface is abstract; hence, new models or even other inversion and DA tools such as hIPPYlib~\cite{VillaPetraGhattas18} can be quickly implemented.
  Fourth, \pyoed leverages development best practices; the entire package is unit tested, a complete API reference is provided, and hands-on examples along with tutorial notebooks are distributed.
  \commentout{
      Fifth, with some effort, \pyoed can, in principle, interface existing OED packages written in Python or R programming languages.  
      \todo{Do we want to say that!}
  }
  
  The rest of this paper is organized as follows. Section~\ref{sec:Background} provides the mathematical formalism of inverse problems, DA, and OED.
  Section~\ref{sec:PyOED} describes the structure and the philosophy of the \pyoed package.
  In Section~\ref{sec:Test_cases} we provide prototypical  experiments and numerical test cases to demonstrate the general workflow and potential of \pyoed.
  Concluding remarks are given in Section~\ref{sec:Conclusions}.

\section{Mathematical Background}
\label{sec:Background}
  In this section we provide a brief overview of the mathematical formulation adopted by \pyoed. 
  Specifically, we describe the forward and the inverse problems in~\ref{subsec:forward_inverse_problem}, and 
  then we introduce the OED formalism in~\ref{subsec:OED}.
  For further details on the mathematical formulation and solution algorithms of OED problems, see, for example,~\cite{attia2022optimal, attia2023robust, alexanderian2021optimal}.

  \subsection{Forward and inverse problem}
  \label{subsec:forward_inverse_problem}
    Simulation models describe the expected evolution of reality, where the \emph{forward problem}
    maps the model parameters (e.g., the initial condition) onto the observation space. 
    A generic definition of the forward problem takes the form
    \begin{equation}\label{eqn:forward_problem_base}
        \obs  = \Fcont(\param) + \obsnoise \,,
    \end{equation}
    where $\param$ is the model parameter of interest, 
    $\obs \in \Rnum^{\Nobs}$ is the observation, and
    $\obsnoise \in \Rnum^{\Nobs}$ is a noise term that accounts 
    for the inaccuracy of the observational system.
    The forward operator $\Fcont$ is occasionally referred to as the ``parameter-to-observable map'' and generally represents a composition of a simulation/solution model $\Sol$ and an observation 
    operator $\ObsOper$.
    The simulation model $\Sol$ describes the evolution of the physical phenomena, 
    for example, space-time advection and diffusion of a contaminant simulated over 
    a predefined model grid.
    The observation operator $\ObsOper$ projects the simulated state onto the observational grid, for example, by interpolation or restriction to the observation grid.
    Thus, the forward problem~\eqref{eqn:forward_problem_base} can be rewritten as 
    $
        \obs  = \ObsOper \circ \Sol(\param) + \obsnoise \,,
    $ where $\circ$ is the composition operator, that is, $\ObsOper \circ \Sol(\param)\equiv \ObsOper \left( \Sol(\param)\right)$.

    Both the simulation model $\Sol$ and the observation operator $\ObsOper$ are imperfect and generally include sensory noise and representativeness errors characterizing imperfection of the map between 
    the model space and observation space.
    The fact that model observations $\Fcont(\iparam)$ are not perfectly aligned with observational data $\obs$ is modeled---assuming additive noise---by adding the noise term $\delta$ to the simulated observations $\Fcont(\iparam)$. 
    In most applications, the observational noise follows a Gaussian distribution
    $\obsnoise \sim \GM{\vec{0}}{\Cobsnoise}$, where $\Cobsnoise$ is the observation error covariance matrix that captures uncertainty stemming from sensory noise and representativeness errors.
    In this case, the data likelihood (needed for the Bayesian formulation of the inverse problem) 
    is
    \begin{equation} \label{eqn:Gaussian_likelihood}
    \CondProb{\obs}{ \param } \propto
      \exp{\left( - \frac{1}{2}
        \sqwnorm{ \Fcont(\param) - \obs }{ \Cobsnoise\inv } \right) } \,,
    \end{equation}
    where the matrix-weighted norm in~\eqref{eqn:Gaussian_likelihood} is defined as
    $\sqwnorm{\vec{x}}{\mat{A}} = \vec{x}\tran \mat{A} \vec{x} $ for a vector $\vec{x}$ and a square symmetric matrix $\mat{A}$ of conformable sizes.

    An \emph{inverse problem} refers to the retrieval of the QoI, that is, 
    the parameter $\param$ from the noisy observation $\obs$, conditioned by the model dynamics.
    In most inverse problem formulations, based on the application of interest, the inversion QoI 
    stated in~\eqref{eqn:forward_problem_base} stands for the calibration parameters of 
    the model, the initial condition, or both. 
    QoI inference can be achieved by finding a point estimate or by building a complete probabilistic description as discussed in Section~\ref{sec:Introduction}. 
    In the former, an optimization problem is solved to minimize the mismatch between the expected observations (through simulation models) and real data. 
    This is typically employed in variational DA methods~\cite{bannister2017review} where the estimate of the true parameter is obtained by minimizing a regularized log-likelihood objective, where regularization is employed to enforce smoothness or background information on the inference parameter.
    In this case, a point estimate of the true $\param$ is obtained
    by solving
    \begin{equation}\label{eqn:fdvar}
      \argmin_{\param} \obj(\param) 
        := \frac{1}{2} \sqwnorm{ \Fcont(\param) - \obs }{ \Cobsnoise\inv }  
         + \frac{1}{2} \sqwnorm{ \param - \paramprior }{ \Cparamprior\inv } \,, 
    \end{equation}
    where $\paramprior$ is an initial guess of the unknown true value of $\param$. 
    In general, the second term is added to enforce regularization or prior knowledge on the solution, for example,
    if the solution is assumed a priori to follow a Gaussian distribution $\param\sim\GM{\paramprior}{\Cparamprior}$. Uncertainty envelopes around the single-point estimate obtained by solving~\eqref{eqn:fdvar} can be developed, for example, by using Laplacian approximation~\cite{tierney1986accurate,stuart2010inverse} where the posterior is approximated 
    by a Gaussian distribution.

    A fully Bayesian approach, on the other hand, aims to provide a consistent probabilistic description of the unknown parameter along with the associated uncertainties and is not limited to Gaussian distributions. 
    This is achieved by describing the posterior, that is, the probability distribution of the model parameter $\param$ conditioned by the available simulations and noisy data $\obs$, and is obtained by applying a form of Bayes' theorem
    \begin{equation}\label{eqn:Bayes}
        \CondProb{\param}{\obs} \propto \CondProb{\obs}{\param} \Prob(\param) \,, 
    \end{equation}
    where $\Prob(\param)$ is the prior, $\CondProb{\obs}{\param}$ is the data likelihood, and $\propto$ indicates removal of a normalizing constant in the right-hand side of~\eqref{eqn:Bayes}.
    For further details on Bayesian inversion see, for example,~\cite{smith2013uncertainty,stuart2010inverse}.
    Given the posterior~\eqref{eqn:Bayes}, one can use the maximum a posteriori (MAP) point as an estimate of the true unknown QoI or follow a Monte Carlo approach to sample the posterior, thus building a complete probabilistic picture; see, for example,~\cite{attia2017reduced}.
    %

    If the forward operator $\Fcont$ is linear (or linearized), and 
    assuming Gaussian observational noise $\GM{\vec{0}}{\Cobsnoise}$ 
    and a Gaussian prior $\GM{\iparb}{\Cparampriormat}$,
    then the posterior is Gaussian $\GM{\ipara}{\Cparampostmat}$ with
    \begin{equation}\label{eqn:Gaussian_Posterior_Params}
    \Cparampostmat = \left(\F \adj \Cobsnoise\inv \F
      + \Cparampriormat\inv \right)\inv \,, \quad
    \ipara = \Cparampostmat \left(
        \F\adj \Cobsnoise\inv \, \obs 
        + 
        \Cparampriormat\inv \iparb
    \right) \,,
    \end{equation}
    where $\F$ is the linear(ized) forward model and $\F\adj$ is the associated adjoint.
    Despite being simple, this setup~\eqref{eqn:Gaussian_Posterior_Params} is of utmost importance in the Bayesian inversion and OED literature and is elementary for testing implementations of new DA and OED approaches, mainly because the posterior can be formulated exactly.
    Moreover, in many large-scale applications, the posterior can be approximated, to an acceptable degree, by a Gaussian distribution obtained by linearizing the nonlinear operator $\Fcont$ around the MAP estimate. 
    The linearized model, also known as the tangent linear model, is  obtained by differentiating $\Fcont$.
    Exploring the posterior in nonlinear inverse problems remains a challenging problem in large-scale settings 
    because of the curse of dimensionality; see, for example,~\cite{vetra2018state, cotter2013mcmc, hairer2014spectral, bardsley2020scalable, beskos2017geometric, attia2017reduced}.
    
    The Bayesian perspective provides a formal mathematical ground for estimating the physical QoI, for example, the model parameter $\iparam$, along with the associated uncertainties given the available sources of information. 
    In many cases, however, this inversion is an intermediate step, and the goal QoI is a function of the model parameter, that is, $\pred :=\PredOper(\param)$.
    A goal-oriented approach is followed in this case where one aims to inspect the posterior of the QoI conditioned by the available data~\cite{LiebermanWillcox13,LiebermanWillcox14}.

  \subsection{Optimal experimental design}
  \label{subsec:OED}
    Here we outline the basics of an OED problem for Bayesian inversion. 
    An excellent review of recent advances in model-constrained OED can be found 
    in~\cite{alexanderian2021optimal}. An OED optimization problem takes the general form
    \begin{equation}\label{eqn:OED_optimization}
        \design\opt = \argmax_{\design} \, \utilityfunc(\design) \,,
    \end{equation}
    where $\utilityfunc$ is a predefined utility function that quantifies the quality of the design $\design$.
    The nature of $\design$ depends on the application at hand, 
    and the utility function $\utilityfunc$ is chosen to enable defining what an ``optimal'' design is. 
    The optimization problem~\eqref{eqn:OED_optimization} is often associated with an auxiliary sparsity-enforcing term $ - \regpenalty \penaltyfunction{\design}$ to prevent dense designs 
    and to reduce the cost associated with deploying observational sensors.
    Thus, the utility function can be be defined as
    $ 
        \utilityfunc(\design) = \Psi(\design) - \regpenalty \penaltyfunction{\design} \,,
    $ 
    where $\Psi(\cdot)$ is an OED optimality objective, referred to hereafter as the ``optimality criterion,'' which is defined based 
    on the inverse problem at hand and on a chosen criterion (e.g., from the well-known OED alphabetic criteria). 
    Design sparsity can be promoted, for example, 
    by setting $\penaltyfunction{\design}:=\wnorm{\design}{0}$. 
    
    Generally speaking, we seek a design that maximizes the utility function $\utilityfunc$. Other formulations, however, involve minimization of an OED optimality criterion; see, for example,~\cite{attia2018goal}. 
    Both formulations are adopted in the OED literature and in \pyoed and often are even equivalent, as explained below.
    In inverse problems, a design can be associated with the observational configuration and thus can be used to optimally select an optimal observational policy. 
    For example, a design can be defined to select sensor location or temporal observation frequency that can help provide accurate prediction with minimum uncertainty, or it can be defined to select 
    an observational configuration that guarantees maximum information gain from the data.

    \paragraph{OED for sensor placement}
      In sensor placement, we associate a binary design variable $\design_i$  with the $i$th candidate sensor location, with $1$ meaning activating the sensor and deactivating it otherwise. 
      This defines the design as a binary vector $\design\equiv\binarydesign\in\{0,1\}^{\Nsens}$, 
      which collectively defines the observational configuration.
      In this case, the OED problem~\eqref{eqn:OED_optimization} takes the form
      \begin{equation}\label{eqn:binary_OED_optimization}
        \design\opt
          = \argmax_{\design \in \{0, 1\}^{\Nsens}} \utilityfunc(\design) := 
          \Psi(\design) - \regpenalty \penaltyfunction{\design}  \,.
      \end{equation}

      In Bayesian OED for sensor placement, the observational configuration is controlled by the observational design. 
      This can be modeled by modifying the observation vector by projection onto 
      the set of active sensors and by modifying the observation error covariance $\Cobsnoise$.
      Specifically, a general formulation follows by replacing $\Cobsnoise\inv$ with a weighted version
      $\wdesignmat(\design)$, resulting in the weighted data likelihood
      \begin{equation}\label{eqn:weighted_joint_likelihood}
        \Like{\obs}{ \param; \design}
          \propto \exp{\left( - \frac{1}{2} \sqwnorm{ \F(\param) - \obs}{\wdesignmat(\design)} \right) } \,,
      \end{equation}
      where the weighted observation error covariance matrix takes the form 
      \begin{equation}\label{eqn:pointwise_weighted_precision}
          \wdesignmat(\design) :=\pseudoinverse{\designmat(\design) \odot \Cobsnoise}\,, \quad
          \designmat_{i,j}(\design)
            :=
            \begin{cases}
              \weightfunc_i\, \weightfunc_j &;\,  i \neq j \\
              \begin{cases}
                0     &; \, \weightfunc_i =0 \\
                \frac{1}{\weightfunc_i^2}  &; \weightfunc_i \neq 0
              \end{cases} &;\, i=j \\
            \end{cases} \,;\,\,
              \begin{matrix}
                & i,j = 1,2,\ldots,\Nsens \,,
              \end{matrix}
      \end{equation}
      where $\odot$ is the Hadamard (Schur) product of matrices,
      $\dagger$ denotes the Moore--Penrose (pseudo) inverse, 
      and $\weightfunc_i\in[0, 1]$ is a weight calculated by using $\design_i$, 
      for example, $\weightfunc_i:=\design_i$; see~\cite{attia2022optimal} for additional details.
      The posterior and the utility function (i.e., the optimality criterion) are then formulated 
      in terms of the modified likelihood.

    \paragraph{The utility function }
      For linear inverse problems, the posterior is Gaussian~\eqref{eqn:Gaussian_Posterior_Params} 
      with modified covariance that depends on the design through the 
      weighted likelihood, that is,
      \begin{equation}\label{eqn:weighted_posterior_cov}
          \Cparampostmat(\design) = \left(\F \adj \wdesignmat(\design) \F
      + \Cparampriormat\inv \right)\inv \,.
      \end{equation}
      However, the posterior uncertainty is independent of the actual realizations of the data.
      This fact enables designing observational policies before actually deploying the observational
      sensors.
      Specifically, in linear Bayesian OED, we set the objective to minimize a scalar summary of the posterior uncertainty, 
      that is, the posterior covariance matrix $\Cparampostmat(\design)$.
      This is the underlying principle of the alphabetical criteria~\cite{pukelsheim2006optimal}.
      For example, an A-optimal design is the one that minimizes the trace of the posterior covariance matrix, 
      and a D-optimal design is the one that minimizes its determinant (or equivalently the log-determinant).
      Note that in the case of a linear model $\F$, the Fisher information matrix $\FIM$ is equal to the inverse of the posterior covariance matrix, that is,
      $
      \FIM(\design) = \Cparampostmat\inv(\design)
        = \F\adj \wdesignmat(\design) \F + \Cparampriormat\inv \,.
      $
      %
      Thus, in this case, the utility function---discarding the penalty term---is set to
      $\utilityfunc(\design):=\Trace{\FIM(\design)}$ for
      A-optimal designs and $\utilityfunc(\design):=\logdet{\FIM(\design)}$ for
      D-optimal designs, and then the utility function is maximized.

      In the case of a nonlinear inverse problem, however, the posterior uncertainty depends on the 
      collected data. For example, the posterior can be approximated by a Gaussian
      centered around the MAP estimate obtained by solving~\eqref{eqn:fdvar}.
      Thus, to obtain an optimal design, one can iterate over finding the MAP estimate of $\iparam$ and solving an OED problem with Gaussian approximation around that estimate.
      Other utility functions employed in nonlinear settings include the information gain as given by the Kullback----Leibler divergence between the posterior and the prior~\cite{huan2014gradient} and its expectation over data known as the expected information gain (EIG) \cite{long2013fast}.

    \paragraph{Popular solution approaches}
      The OED problem~\eqref{eqn:binary_OED_optimization} can be viewed as a mixed-integer 
      program and can be solved by using branch-and-bound~\cite{Gerdts05,Leyffer01}. 
      This type of research, however, has not yet been applied to model-constrained OED.
      A common approach to solving~\eqref{eqn:binary_OED_optimization} is to replace the binary optimization with the following relaxation:
      \begin{equation}\label{eqn:relaxed_oed}
        \design\opt
          = \argmax_{\design \in [0, 1]^{\Nsens}} \,
        \utilityfunc(\design) \,,
      \end{equation}
      where the design variables are relaxed to take any values in the interval $[0, 1]$ rather than only the binary values $\{0, 1\}$. This approach, if carried out properly, has the effect of generating a continuous relaxation surface that connects the values of the objective evaluated at the binary designs; see~\cite{attia2022optimal} for further details. 
      A gradient-based optimization approach is then followed to solve~\eqref{eqn:relaxed_oed}. 
      Gradient formulation, however, is mathematically involved and can be extremely computationally demanding because it requires numerous evaluations of the forward operator and the corresponding adjoint.
      Moreover, this requires both the optimality criterion $\Psi$ 
      and the penalty $\penaltyfunc$ to be differentiable (or be replaced with differentiable
      approximations)
      with respect to the design $\design$.
      The computational challenge can be ameliorated by employing inexpensive surrogates to approximate
      the forward operator $\Fcont$ and the corresponding adjoint; see, for example,~\cite{oleary2022derivative, oleary2022learning, wu2023large}.

      A stochastic learning approach to binary OED has been recently presented in~\cite{attia2022stochastic} 
      to solve the binary optimization problem~\eqref{eqn:binary_OED_optimization} without 
      the need for relaxation.
      This approach does not require differentiability of the utility function $\utilityfunc$. 
      In this approach, the optimal design is defined as
      \begin{equation}\label{eqn:stochastic_OED_optimization}
        \hyperparam\opt
          = \argmax_{\hyperparam\in[0,1]^{\Nsens}}
              \Expect{\design\sim\CondProb{\design}{\hyperparam}}{\utilityfunc(\design) - \baseline} \,, 
      \end{equation}
      where $\CondProb{\design}{\hyperparam}$ is a multivariate Bernoulli distribution with parameter
      $\hyperparam$ specifying probabilities of success/activation of each entry of $\design$, that is, $\hyperparam_i\in[0, 1]$.
      Here, $\baseline$ is a constant ``baseline'' used to minimize variability of the stochastic estimate of the gradient; see~\cite{attia2022stochastic} for further details.
      
      Bayesian inversion and model-constrained OED are increasingly interesting research topics.
      \pyoed aims to be a platform that can help explain the details of existing formulations 
      and algorithms, to enable rapid development of new ideas in one or more of the components 
      of an OED problem, and to enable comparing new methods with existing approaches.
      For example,~\pyoed has been recently employed to develop new approaches for robust
      OED~\cite{attia2023robust}. 

\section{\pyoed: Structure and Philosophy}
\label{sec:PyOED}
\begin{figure}[!htbp]
    \centering
    \includegraphics[width=0.75\textwidth]{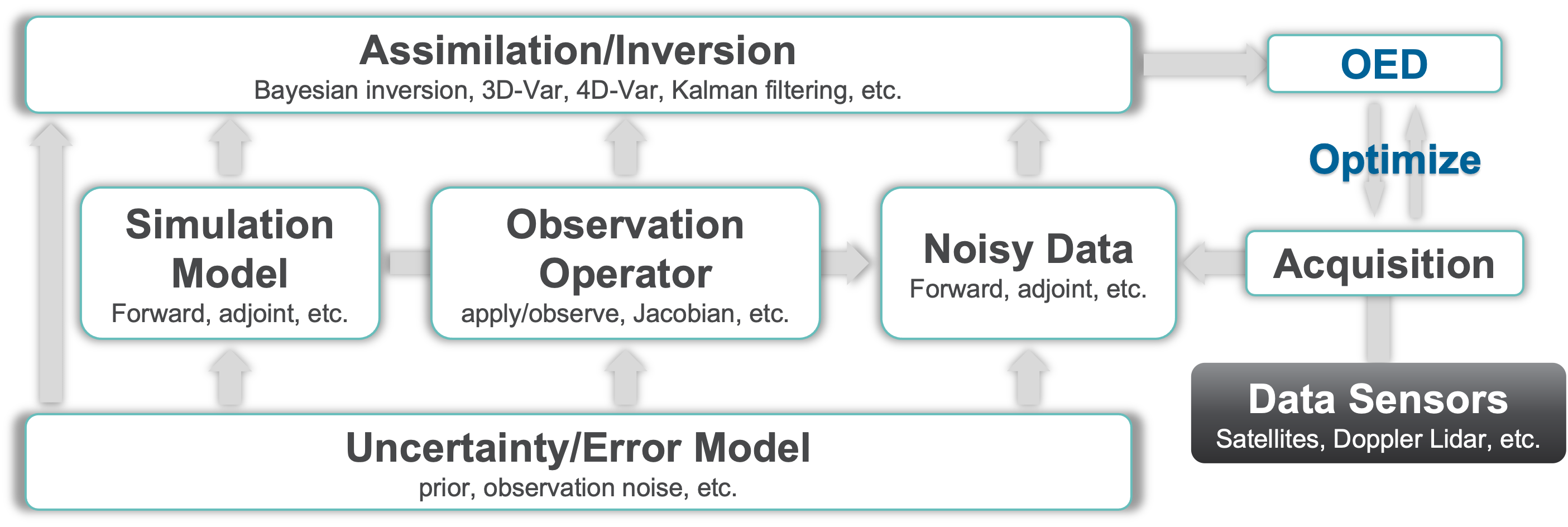}
    \caption{Main components of a PDE-constrained OED system.}
    \label{fig:DA_and_OED}
  \end{figure}

  %
  Solving model-constrained OED and inverse problems requires proper understanding and formulation of the underlying dynamical system, observational configuration, uncertainty models, DA and inversion algorithms, and the OED objective and the selected utility function~\cite{ucinski2000optimal}. 
  The main components of a PDE-constrained OED system are sketched in Figure~\ref{fig:DA_and_OED}.
  \pyoed is a stand-alone, yet extensible, Python package that  provides users and researchers in the computational science and engineering disciplines with a testing suite that effectively implements and enables interfacing these components in an OOP fashion. 
  For example, \pyoed provides a variety of time-dependent and time-independent simulation models. 
  These include systems governed by linear algebraic equations, ordinary differential equations, and PDEs. 
  \pyoed is also equipped with 
  a set of classes implementing various observational operators,
  probabilistic uncertainty and error models, 
  variational and Bayesian DA algorithms, 
  novel optimization methods, and various OED approaches. 
  A high-level overview of the \pyoed major components and their coupling for solving DA and OED problems is provided in Figure~\ref{fig:DA_and_OED}.
  The rest of Section~\ref{sec:PyOED} describes the main components of \pyoed and outlines the functionality they provide in correspondence with Figure~\ref{fig:DA_and_OED}.

  \subsection{Code structure}
  \label{subsec:code_structure}
\begin{figure}[!htbp]
      \centering
      \includegraphics[width=0.85\textwidth]{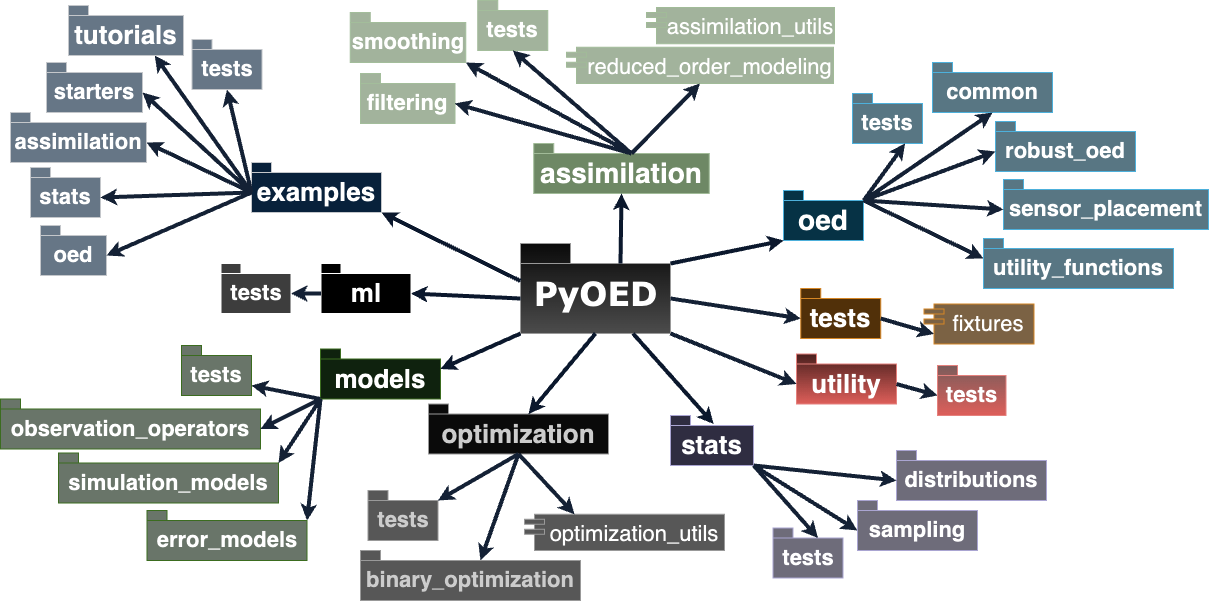}
        \caption{Main subpackages available in the current version of \pyoed (\pyoedversion).}
      \label{fig:PyOED_Subpackages}
    \end{figure}

    Figure~\ref{fig:PyOED_Subpackages} shows the main subpackages shipped with the core of \pyoed.
    The rest of this section
    provides a high-level description, in alphabetic order, 
    of the packages/subpackages of~\pyoed as displayed in Figure~\ref{fig:PyOED_Subpackages}.
    %

    \paragraph{\code{pyoed.assimilation}}
      \pyoed provides a set of DA tools that include algorithms for ``filtering'' and ``smoothing.''
      These two terms are widely used in the DA literature as two distinctive categories of 
      inverse problems and solution algorithms~\cite{attia2015hmcfilter,attia2015hmcsmoother}. 
      A filtering algorithm solves inverse problems that involve time-independent or time-dependent 
      simulation models, while smoothing algorithms are restricted to time-dependent models.
      Filtering involves prediction (of observation) using the parameter-to-observable map, 
      followed by a correction procedure to correct knowledge of the QoI given the observational data.
      In filtering for time-dependent simulations, the observational data is assimilated sequentially, 
      with one observation time per assimilation window/cycle. 
      Examples of filtering DA methods include three-dimensional variational DA and Kalman filtering~\cite{evensen2009data,asch2016data}. 
      Smoothing, on the other hand, is concerned with history matching; 
      these algorithms try to find the QoI that best matches multiple spatiotemporal
      observations (a trajectory), usually defined as an initial value problem.
      Examples include space-time Bayesian inversion~\cite{stuart2010inverse}
      and four-dimensional variational DA~\cite{asch2016data}. 
      Solving an inverse problem generally requires multiple evaluations of the simulation model,
      which can be computationally intensive. This situation can be partially ameliorated by utilizing 
      a surrogate or a reduced-order model (ROM)~\cite{attia2017reduced,wu2023large} to approximate 
      the simulation model effect.
      Implementations of filtering DA algorithms are provided through \code{pyoed.assimilation.filtering}, and 
      smoothing algorithms are provided in \code{pyoed.assimilation.smoothing}. 
      Basic ROM implementations are also provided through \code{pyoed.assimilation.reduced\_order\_modeling}.
      
      Note that all the  DA implementations in \pyoed are equipped with tools to handle both linear and nonlinear simulation and observation models.
      In the linear case the DA methods in \pyoed provide a Gaussian representation of the posterior, which has an analytical form~\eqref{eqn:Gaussian_Posterior_Params}. 
      In the case of a nonlinear process, however, one can approximate the true posterior via a Laplace approximation approach, for example by constructing a Gaussian centered around the solution of the optimization problem~\eqref{eqn:fdvar}. 
      \pyoed filtering and smoothing algorithms, both variational such as 3DVar and 4DVar and Bayesian such as Kalman filtering, follow a Laplacian approach to model the posterior.
      This enables researchers to explore the structure of the posterior by simply calling a \code{sample} method associated with the posterior. 
      For full exploration of the exact posterior, one would need to utilize general sampling methods, such as Markov chain Monte Carlo (MCMC), which are provided 
      through the \code{stats} subpackage as discussed below.
      
    \paragraph{\code{pyoed.examples}}
      This subpackage provides various example scripts that users can follow 
      to learn how to effectively use different pieces of the package. 
      The modules in this subpackage explain how to load all pieces of 
      the package independently and describe how to properly coordinate 
      these components to design a consistent DA and/or OED experiment. 
      
      Given the popularity of Jupyter Notebooks in the computational science community, 
      we converted some of the examples in the subpackage \code{pyoed.examples} 
      to Jupyter Notebooks and provided them in \code{pyoed.examples.tutorials}.
      We also employ some of them in the test cases presented in Section~\ref{sec:Test_cases}. 
      A growing list of tutorials can be found on the \pyoed documentations 
      web page~\cite{pyoedwebpage2023}.
    
    \paragraph{\code{pyoed.ml}}
      This subpackage is intended to provide implementations of machine learning 
      algorithms useful for DA and OED applications.
      For example, the stochastic learning approach to 
      OED~\eqref{eqn:stochastic_OED_optimization} can be seen as 
      a reinforcement learning (RL) approach to solving the OED problem.
      The module \code{pyoed.ml.reinforcement\_learning} under this package provides 
      implementations of the main RL components, including an agent, 
      a policy, transition probability, actions, and utility functions.

    \paragraph{\code{pyoed.models}}
      Following the convention in the \dates package~\cite{attia2019dates}, we use the 
      word ``model'' to refer to three entities: 
      \emph{the simulation model}, 
      \emph{the observation model (or operator)}, and 
      \emph{the error models}.
      
      The simulation model provides a prediction about the behavioral pattern of the physical 
      phenomena of concern.
      In \pyoed, we differentiate two types of simulation models: time-independent and time-
      dependent simulations. Specifically, \pyoed provides various simulation models under the 
      \code{pyoed.models.simulation\_models}, including several versions of the Lorenz 
      system~\cite{lorenz1996predictability}, and advection-diffusion models. 
      The structure of these prototypical simulation models should provide clear guidelines 
      to practitioners willing to adopt \pyoed for their particular applications. 
      
      The observation operator maps the model state onto the observation grid, thus providing 
      a functional mapping between the model state and observational data. 
      Two of the most prominent observation operators in experimental settings are 
      the identity operator and an interpolator.
      \pyoed provides implementations of several observation operators including these two, 
      with an observational design properly incorporated to enable altering observational 
      configurations at any point in the DA or OED solution process. 
      Observation operators are provided in the \code{pyoed.models.observation\_operators} subpackage. 

      The error models quantify the uncertainty associated with the model parameter, model state, 
      and observational data.
      An experimental design can be associated with any of these pieces. 
      For example, in sensor placement, an experimental design is associated with the 
      observational grid; thus, modifying the observational design affects the observational 
      error model. 
      For example, in the relaxation approach~\eqref{eqn:relaxed_oed} the design weights 
      scale the entries of the covariance (or the precision) matrix, 
      and the stochastic approach~\eqref{eqn:stochastic_OED_optimization} works by removing 
      rows/columns of the observation error covariance matrix corresponding to zero 
      design variables.
      \pyoed provides various implementations of error models suitable for modeling priors, 
      as well as model prediction and observational errors in Bayesian inversion, where a design 
      variable is consistently implemented to enable modifying the experimental design during any 
      step of the DA and/or OED solution process.
      \pyoed provides various error model implementations through the subpackage 
      \code{pyoed.models.error\_models}, including multiple versions of Gaussian and 
      Laplacian models suitable for a wide range of applications and settings. 
      
    \paragraph{\code{pyoed.oed}}
      OED is the main component of \pyoed that provides implementations of various algorithmic
      approaches for solving OED problems, including relaxation~\eqref{eqn:relaxed_oed} and 
      stochastic learning~\eqref{eqn:stochastic_OED_optimization}, 
      as well as recent developments including robust OED~\cite{attia2023robust}.
      Most implementations in this package take an inverse problem (DA object) as input 
      and use it to access all the underlying components, thus gaining access to 
      the simulation model, error models, and observation operator as well
      as the experimental design.
      This approach enables the user to modify an experimental design, solve the DA problem if needed, 
      and solve the underlying OED optimization problem.
      Each of the OED implementations in \pyoed returns an object that provides access to the OED problem itself, 
      results of the OED solution process, and functionality to visualize the results.
      
      Utility functions (optimality criteria) are the cornerstone of OED problems 
      since they define how the design affects the inverse problem and whether the OED optimization problem 
      is a maximization (e.g., trace of the FIM) or a minimization (e.g., trace of the posterior covariance matrix). 
      \pyoed provides various utility functions and optimality criteria  through~\code{pyoed.oed.utility\_functions}. 
      For example, common alphabetic criteria such as A- and D-optimality 
      are provided in~\code{pyoed.oed.utility\_functions.alphabetic\_criteria}; 
      and common information-theoretic utility functions, such as the EIG, 
      are provided in~\code{pyoed.oed.utility\_functions.information\_criteria}. 
      These utility functions' implementations are general-purpose as they are designed to be agnostic 
      to the way the design enters the inverse problem.
      Specifically, they take as input (upon instantiation) an OED problem object and use it to modify 
      the design in order to evaluate the value of the utility function and optionally its derivative. 
      The OED problem object itself is responsible for defining how the design modifies the inverse problem.
      %
      Specialized utility functions for sensor placement are provided through~\code{pyoed.oed.utility\_functions.sensor\_placement}.
      These implementations extend the generic alphabetic and information criteria implementations discussed above,
      to enable modifying the way the design affects the observational configurations such as the observation error model.

      While the alphabetic criteria are ideal for Bayesian OED for linear inverse problems,
      OED for nonlinear inverse problems remains an active research topic.
      Utility functions valid for nonlinear problems include 
      employing a Laplace (Gaussian) approximation to the posterior
      by linearization of the parameter-to-observable map, 
      employing KL divergence between the prior and the posterior, 
      and general information-theoretic utility functions such as 
      the EIG.
      
      All utility functions available through \code{pyoed.oed.utility\_functions} 
      are capable of handling both linear and nonlinear inverse problems; however,
      the way nonlinearity is handled depends on each implementation. 
      For example, the EIG is independent of the linearity of the problem.
      The alphabetic (A, D, etc.) criteria, on the other hand, assume a Gaussian 
      posterior and thus follow a Laplace approximation approach
      by linearizing the parameter-to-observable map, for example, around the solution of the inverse problem.
      The choice of the linearization point controls the quality of the linearization 
      and is thus set to be a configuration parameter. 
      Default values of configuration parameters in all \pyoed classes are provided to enable using the code with minimal 
      effort.
      
    \paragraph{\code{pyoed.optimization}}
      Numerical optimization routines are elementary for solving OED optimization problems,
      as well as variational DA algorithms.
      A variety of optimization software packages can be used for solving 
      numerical optimization problems including those described in this work.
      \pyoed enables using external optimization packages, including  
      Python's \code{Scipy} package, to solve DA and OED optimization problems.
      \pyoed, however, provides specific implementations of optimization procedures
      not available in popular optimization packages, such as the stochastic algorithm
      described in~\cite{attia2022stochastic} 
      and robust binary optimization~\cite{attia2023robust}.  
      
    \paragraph{\code{pyoed.stats}}
      This subpackage aims to collect statistical procedures used by other parts of the package, 
      such as sampling routines, and distribution classes implementing random variables and their probabilistic 
      utility functions including density evaluation and log probabilities.
      For example, \pyoed provides an exemplary implementation of a multivariate Bernoulli distribution required by
      the RL algorithms in \code{pyoed.ml.reinforcement\_learning}.
      Sampling routines including rejection sampling and various MCMC implementations including 
      Hamiltonian Monte Carlo  samplers~\cite{attia2015hmcfilter,attia2015hmcsmoother} are provided through~\code{pyoed.stats.sampling}.
      This subpackage also provides access to multiple proposals that can be employed in MCMC samplers.
      Since statistical tools are crucial for various DA and OED algorithms, we chose to keep the subpackage \code{pyoed.stats} rather 
      than moving these implementations to other parts of the package. 
      This approach is advantageous because we continuously extend the package with various statistical tools, for example, 
      for randomized approximation methods for Bayesian inversion.

    \paragraph{\code{pyoed.tests}}
      \pyoed follows best practices by providing a consistent set of unit tests to 
      assure the quality of the package as it continues to grow.
      This subpackge provides general fixtures and various testing modules and functions used by all unit tests across the package.
      Currently we  put all unit tests of each subpackage under the corresponding subpackage directory, 
      however, this approach can change in the future without impacting the package itself or causing any backward compatibility issues.
      Note that in general in \pyoed random number generation is managed properly through simplified, yet powerful, mixin classes 
      developed in~\code{pyoed.utility.mixins}. These are imported by any class that performs any random number generation.
      Random number generation is controlled by a random seed chosen upon instantiation of an object. 
      This enables proper reproducibility of all test cases provided in the package and also enables robust unit 
      testing as random seeds are properly set for all unit tests.
      %

    \paragraph{\code{pyoed.utility}}
      This subpackage aims to collect implementations of general-purpose 
      functionality, such as file I/O and visualization, as well as
      general mathematical and statistical procedures. 
      The subpackage includes various visualization tools and 
      matrix-free implementations of expensive operations
      such as evaluating the trace and log-determinant of a matrix. 
      It also provides routines to approximate matrix trace and determinant
      (or its logarithm) using statistical randomization~\cite{AlexanderianSaibaba17,han2015large,saibaba2016randomized}.

    While the components of DA and OED problems discussed above can be used independently of each other, 
    some level of ordering is mandatory for proper utilization.
    For example, an inverse problem (DA) object cannot be instantiated before a simulation model, 
    an observation operator, and error model objects. 
    Similarly, an OED problem for Bayesian inversion cannot be solved before creating an inverse problem.
    A practical guide that illustrates how to follow this simple workflow is described in Section~\ref{sec:Test_cases}.

  \subsection{Extending and contributing to \pyoed}
  \label{subsec:extending_PyOED}
    Although the package contains many common models, assimilation processes, and OED algorithms, 
    we recognize that many users will want to replace individual parts of the workflow.
    Hence, \pyoed's core is designed to be easily extended.
    Each individual component corresponds to a common interface for that layer and inherits a proper abstract base class. 
    For example, every simulation model inherits the \code{pyoed.models.simulation\_model.SimulationModel} abstract base class. 
    The internal details of how the specified abstract methods are implemented are black-boxed.
    Thus, \pyoed leaves a user free to reimplement components as long as they fulfill the corresponding abstract base class.
    We have personally used the library in conjunction with others such as hIPPYlib~\cite{VillaPetraGhattas18}, for solving high-dimensional inverse problems governed by PDEs, and Jax~\cite{james2018jax}, for automatically differentiating through an ordinary differential equation.
    Guided by this philosophy, \pyoed itself is kept fairly lean in terms of dependencies, as each additional external package adds another dimension of complexity to maintenance.
    This can be viewed as a form of defensive software design against external deprecation, to minimize internal API breaking changes down the road.
    To that end, extensions and contributions to \pyoed that involve additional dependencies will likely be done through a plug-in ecosystem, as opposed to being merged into the core.
    Plans for such a ecosystem are future work.
    
  \subsection{Code availability}
  \label{subsec:code_availability}
    Online documentation of \pyoed is available through 
    \cite{pyoedwebpage2023},
    and the 
    development version of the package is available from 
    \cite{pyoedrepo2023}. 
    A growing list of examples and tutorials is available from~\cite{pyoedrepo2023}.
    Moreover, detailed description of requirements, 
    installation instructions, and  proper unit testing of the package
    is available through~\cite{pyoedrepo2023,pyoedwebpage2023}.

  \section{Test Cases}
  \label{sec:Test_cases}
    \commentout{
        Referee 3 has the following concern. Make sure it is properly handled below.
        Section 4.1. 
        What exactly is the goal of this section, i.e., what is the problem that is being targeted? 
        Is it an inverse problem or an OED problem governed by an inverse problem constrained by the time-dependent forward problem?
        If the former, what is the inversion parameter/variable? 
        Or for OED, what OED criterion is being applied and how is the problem solved?
        Section 4.2. Similar to the question for Section 4.1., what is the inversion parameter? what is the OED problem?
    }

    \pyoed comes with a set of numerical examples of increasing complexity, 
    so that users can gradually get familiar with the machinery of \pyoed. 
    The complete list of available examples can be found under \code{pyoed.examples}, with 
    the Jupyter Notebook tutorial demos  located under \code{pyoed.examples.tutorials}.
    For the sake of illustration, we briefly highlight two of those tutorials.
    
    The first test case, discussed in Section~\ref{subsec:ToyLinear}, corresponds to 
    the notebook~\code{pyoed.examples.tutorials.toy\_linear}.
    This tutorial employs a simple one-dimensional toy model characterized by a linear Gaussian setup to explain 
    the workflow that should be generally followed by the users. It aims to clearly explain
    how all elements of the DA and OED processes in \pyoed can be instantiated and  utilized.
    %
    The second test case, presented in Section~\ref{subsec:results:standard_oed_examples}, corresponds to 
    the notebook~\code{pyoed.examples.tutorials.OED\_AD\_FE}. 
    This test case employs a two-dimensional advection-diffusion problem defining the transport of contaminant in a given environment.
    This setup has served as a standard experiment widely used in OED scientific research; 
    hence we discuss the details of its implementation in Section~\ref{subsec:results:standard_oed_examples}.

  \subsection{An ideal setup: linear Gaussian toy problem}
  \label{subsec:ToyLinear}
    In this test case, both the simulation model and the observation operator are linear, 
    and the error models (observation noise and prior) are Gaussian, and thus the posterior is also Gaussian with 
    moments (mean and covariance) having closed forms~\eqref{eqn:Gaussian_Posterior_Params}.
    Such a simplified setup is useful for testing new formulations in both DA and OED, and thus it is provided in \pyoed.
    
    \paragraph{The forward and inverse problems}
    In this test case, we consider a time-dependent forward model defined at time 
    instances $t_0 + i \Delta t$ for $i = 0, \dots, \nobstimes$ with fixed time step size $\Delta t > 0$.
    The time evolution of the state $\modelstate$ over time instances is modeled by a matrix vector product and is given by the expression
    \begin{align} \label{eqn:Toy_Linear_Model}
        \modelstate_n = \mat{A} \modelstate_{n-1}, 
        \qquad
        \obs_n = \mat{I} \modelstate_n + \obsnoise,
        \qquad 
        n=1, 2, \ldots \,,
    \end{align}
    where $\modelstate_{n} \in \Rnum^{\Nstate}$ is the discrete model state at time 
    instance $t_n$ and $\mat{A} \in \Rnum^{\Nstate \times \Nstate}$ is 
    a square matrix representing model evolution over time interval $[t_{n-1}, t_{n}]$. 
    We impose an identity observation operator $\mat{I}$ to determine $\obs_n$ 
    from $\modelstate_n$, along with additive Gaussian observation error $\obsnoise \sim \GM{\vec{0}}{\mat{R}}$.
    Additionally, we assume a Gaussian prior on the initial state, that is, $\modelstate_0 \sim \GM{\modelstate_0^{\rm pr}}{\mat{\Gamma}_{\rm pr}}$.
    Together, these assumptions imply that the posterior is Gaussian $\GM{\modelstate_{0}^{\rm post}}{\mat{\Gamma}_{\rm post}}$ with 
    \begin{equation}\label{eqn:Gaussian_Posterior_Toy}
      \mat{\Gamma}_{\rm post} = \left(
        \mat{A}\tran \mat{R}^{-1} \mat{A} + \mat{\Gamma}_{\rm pr}^{-1} 
      \right)^{-1} \,,
      \qquad
      \modelstate_{0}^{\rm post} = \mat{\Gamma}_{\rm post} 
      \left( 
        \mat{\Gamma}_{\rm pr}^{-1} \modelstate_{0}^{\rm pr}
        + \sum_{i=1}^{\nobstimes}{ \mat{A}\tran \mat{R}^{-1} \obs }
      \right) \,.
    \end{equation}
    Since~\eqref{eqn:Gaussian_Posterior_Toy} is a closed form of the posterior, 
    we can use it to test and debug new DA and OED implementations. 
    This fact is highly utilized in the unit tests developed in \pyoed. 
    This specific problem can be found in the Jupyter Notebook in \code{pyoed.examples.tutorials.toy\_linear} 
    and can be used to regenerate the following numerical results.
    In particular, we discuss solving the inverse problem~\eqref{eqn:Gaussian_Posterior_Toy} to identify the true value 
    of the inference parameter $\modelstate_{0}$.

    Let us select $\Nstate = 5$ and create a \code{pyoed.models.simulation\_models.toy\_linear.ToyLinearTimeDependent}
    model with a randomly generated matrix $\mat{A} \in \Rnum^{\Nstate \times \Nstate}$. 
    Once the \code{model} object is instantiated, one can call the method \code{get\_model\_array()} to generate a matrix representation 
    of the model (for this linear case) and can employ  a method \code{create\_initial\_condition()} to create a ground truth for 
    the inference parameter $\modelstate_{0}\true$.
    With a default random seed set for the model in this test case (as discussed earlier), the model array and the ground truth are given by the following:
    \begin{equation}
        \mat{A} := \begin{bmatrix}
            -0.99  & -0.37  &  1.29  &  0.19  &  0.92   \\
             0.58  & -0.64  &  0.54  & -0.32  & -0.32   \\
             0.10  & -1.53  &  1.19  & -0.67  &  1.00   \\
             0.14  &  1.53  & -0.66  & -0.31  &  0.34   \\
            -2.21  &  0.83  &  1.54  &  1.13  &  0.75 
        \end{bmatrix} \,; \qquad
        \modelstate_{0}\true := 
            \begin{bmatrix}
                  -0.36  \\ -1.23   \\ 1.23  \\ -2.17 \\-0.37
            \end{bmatrix} \,.
    \end{equation}

    Note that this model simulates a time-dependent phenomenon where applying $\mat{A}$ to a state vector represents simulating 
    the evolution of the phenomenon over an interval $[0, t]$ with a model time step in this example set (in the model configurations) to $t=0.1$.
    We select a standard multivariate Gaussian prior $\GM{\vec{\modelstate_{0}\pr}}{\mat{\Cparamprior}}$ with 
    $\modelstate_{0}\pr=\vec{0}, \,\Cparamprior=\mat{I}$, 
    and we impose an observation error variance of 
    $0.001$, that is, $\Cobsnoise = 10^{-3} \mat{I}$, with $\mat{I}\in\Rnum^{5\times5}$ being the identity matrix.
    Then, using the relation defined in~\eqref{eqn:Toy_Linear_Model}, we generate synthetic data in the time window $[0, 0.3]$ by adding
    realizations of the noise errors, drawn from the observation error model, to the true trajectory obtained by simulating~\eqref{eqn:Toy_Linear_Model} 
    with $\modelstate_{0}= \modelstate_{0}\true$.
    Specifically, this examples sets $\Delta t=0.1$ and generates synthetic observations $\obs$ at $\nobstimes=3$ time instances $t \in\{0.1, 0.2, 0.3\}$. 

    A \code{pyoed.assimilation.smoothing.fourDVar.VanillaFourDVar} DA object is then created. 
    By registering all the elements above into the DA object, we can  solve the inverse problem and assess its performance against the ground truth given by~\eqref{eqn:Gaussian_Posterior_Toy}.
    In this case the DA object solves following optimization problem:
    \begin{equation}\label{eqn:toy_linear_fdvar}
      \modelstate\opt = \argmin_{\modelstate} \obj(\modelstate) 
        := \frac{1}{2} \sum_{i=1}^{3} \sqwnorm{ \mat{A}^{i} \modelstate_{0} - \obs_i }{ \Cobsnoise\inv }  
         + \frac{1}{2} \sqwnorm{ \modelstate_{0} - \modelstate_{0}\pr }{ \Cparamprior\inv } \,, 
    \end{equation}
    where in this case $\modelstate_{k\,t}=\mat{A}\modelstate_{0}=\mat{A}^{k}\modelstate_{0}$ with $t=0.1$ the model time step.
    The optimization procedure chosen to solve the inference problem~\eqref{eqn:toy_linear_fdvar} is a configuration parameter 
    that can be chosen upon instantiation of the DA object and can be altered at any point by calling the right registration method,
    in this case, \code{register\_optimization\_routine()}, and by passing the name of the optimizer to be used. 
    The optimization routine used here is L-BFGS-B provided by the \code{scipy.optimize} package. 
    Another variant of the code in this tutorial is given in~\code{pyoed.examples.starters.toy\_linear} with more details. 
    Additionally, among other scripts, a simplified nonlinear inverse problem utilizing a nonlinear observation operator 
    is provided in~\code{pyoed.examples.starters.toy\_linear}.

\begin{figure}[!htbp]
      \centering
        \includegraphics[width=0.45\textwidth]{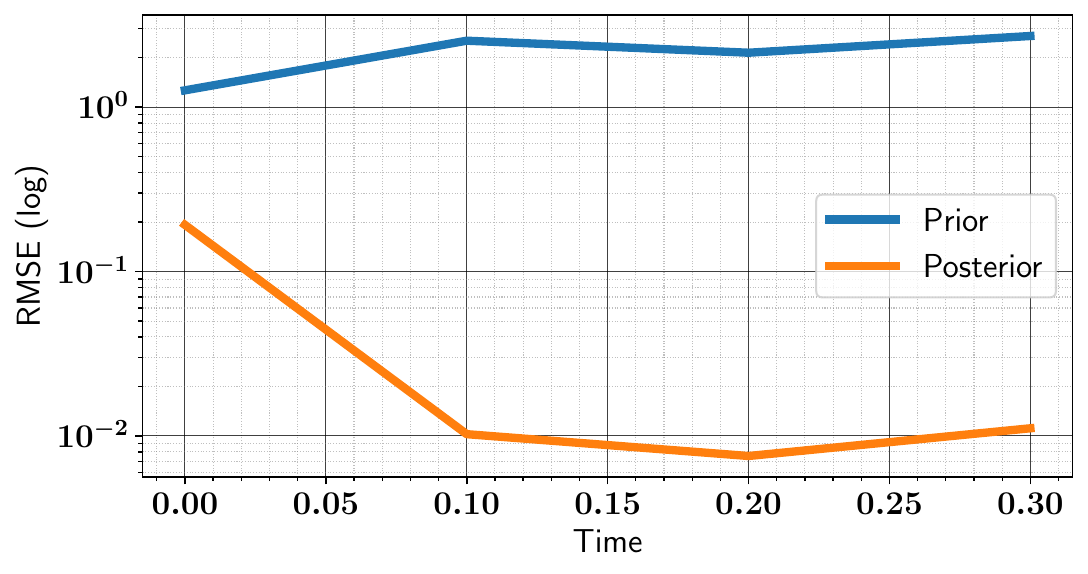}
        \caption{RMSE results of the solution of the inverse problem presented in Section~\ref{subsec:ToyLinear}.
        A ground truth (true trajectory) is found by integrating the 
        simulation model~\eqref{eqn:Toy_Linear_Model} with a predefined 
        true initial condition $\modelstate_0$, over the simulation timespan 
        $[t0, t_0 + \nobstimes \Delta t]$. 
        The prior RMSE is obtained by calculating the average (over all time instances) RMSE between the prior trajectory and the ground truth. 
        The prior trajectory is simulated with the initial condition $\modelstate_0$ set to the prior mean.
        Similarly, the posterior RMSE is found by comparing the trajectory simulated by setting $\modelstate_0$ to the posterior mean with ground truth.
        }
      \label{fig:toy_linear_rmse_traject}
    \end{figure}

    The method \code{solve\_inverse\_problem()} provided by the DA object can be invoked to solve 
    the optimization problem~\eqref{eqn:toy_linear_fdvar} and return a single-point estimate of $\modelstate_{0}\true$.
    \pyoed provides several functions to evaluate the quality of the solution of the optimization problem, such as 
    the root mean squared error (RMSE). 
    For instance, we use the function \code{pyoed.utility.calculate\_rmse}  to inspect the RMSE results for the prior and the analysis (posterior) means over the whole assimilation timespan, with results plotted in Figure~\ref{fig:toy_linear_rmse_traject}. 
    One can clearly see that the RMSE for the prior mean monotonically increases with time, while the RMSE for the posterior remains at significantly smaller values across the assimilation window, which speaks to the quality of the inverse problem algorithm.

\begin{figure}[!htbp]
      \centering
        \includegraphics[width=0.65\textwidth]{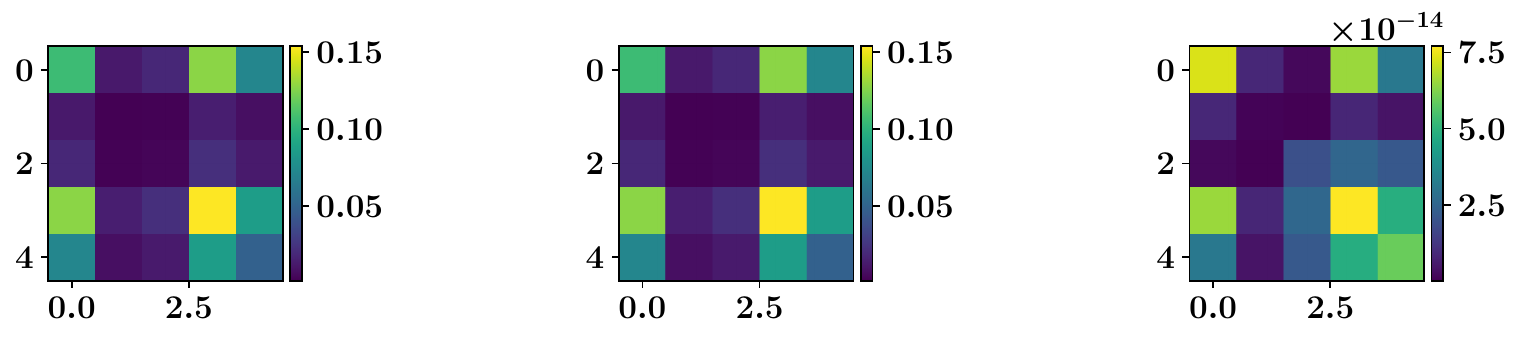}
        \caption{Entries of the posterior covariance matrix and the associated errors.
          Left: the posterior covariance matrix generated by \code{inverse\_problem.posterior.covariance\_matrix()} functionality.
          Middle: the closed-form posterior covariance matrix given by~\eqref{eqn:Gaussian_Posterior_Toy}.
          Right: RMSE obtained by pointwise comparison of the covariance matrices obtained by solving 
          the inverse problem (left) and by using the closed form (middle).}
      \label{fig:toy_linear_posterior_covariance}
    \end{figure}

    All DA methods in \pyoed provide access to a \code{posterior} object (error model) that models 
    the Bayesian posterior of the inverse problem or approximates it.
    Thus, one can also analyze the quality of the assimilated posterior by generating and plotting the posterior covariance matrix against the closed-form expression in~\eqref{eqn:Gaussian_Posterior_Toy}.
    While this is infeasible for high-dimensional models, it is perfectly tractable for this toy model.
    We visualize this comparison along with the mismatch errors in Figure~\ref{fig:toy_linear_posterior_covariance}.
    Again, one can see that there is an agreement between the posterior covariance matrix computed by using both methods, with RMSE levels close to machine precision.

\begin{figure}[!htbp]
      \centering
        \includegraphics[width=0.65\textwidth]{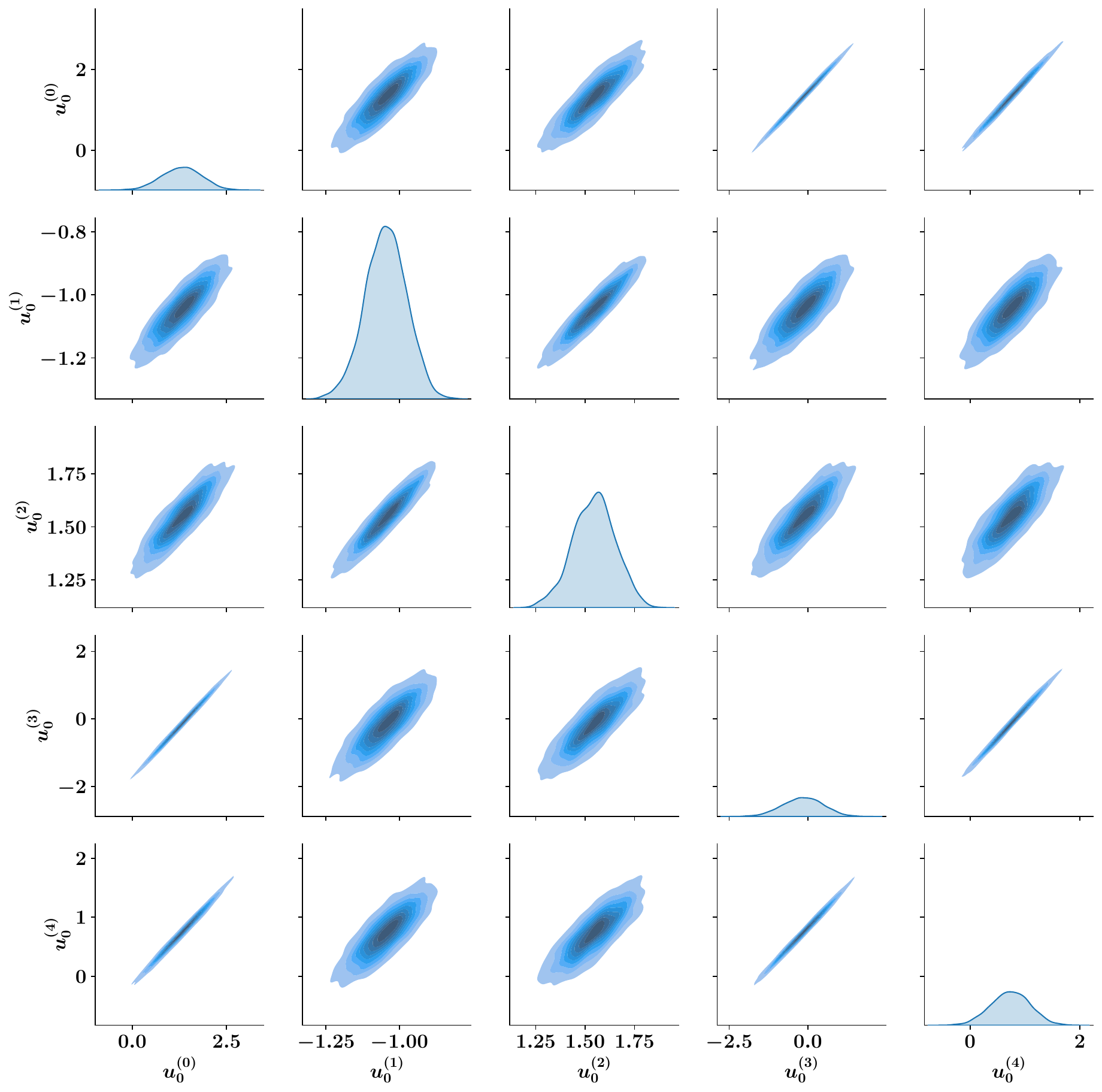}
        \caption{
            Pairplot of $2,000$ samples generated from the posterior~\eqref{eqn:Gaussian_Posterior_Toy}.
        }
      \label{fig:toy_linear_posterior_samples}
    \end{figure}

    One would also be interested in solving the inverse problem corresponding to the observation configuration 
    determined by the optimal design (or any other design). 
    \pyoed provides an easy-to-use interface to fulfill such a requirement. 
    Specifically, each OED object has a \code{design} attribute that enables proper retrieval and update of the experimental design.
    One can use it to set the experimental design to the optimal design provided through the \code{OEDResults} object and then just
    call the \code{solve\_inverse\_problem()} method of the DA object.
    Once the inverse problem is solved, one can test the posterior, for example, to generate samples from the posterior.
    The easiest approach is to call the \code{sample()} method provided by all posterior objects (and all error models), but one also can employ MCMC sampling methods as discussed above.
    In Figure~\ref{fig:toy_linear_posterior_samples} we show a pairplot of the samples generated by invoking the \code{posterior.sample()} method associated with the \code{posterior} accessible through the DA object.
    %
    
    \paragraph{The OED problem}
    After building an inverse problem (here a 4DVar object), we can easily perform OED.
    Since the observation space in this example has five degrees of freedom, we can consider a sensor placement model where there are five candidate sensor locations for each observation gridpoint, and we wish to discover the best pair of sensors that  maximizes a chosen utility function.
    \pyoed provides many different implementations of OED algorithms and utility functions so that users can build their own  solver for their situation.
    In this case, we will evaluate fitness using the D-optimal design criterion with a penalty function $\Phi(\zeta) = (|\zeta| - 2)^2$ with weight $-20$ 
    as described in Section~\ref{subsec:OED}.
    To optimize, we will use a relaxed OED formulation with a pointwise weighted covariance.
    Specifically, the OED optimization problem here takes the form
    \begin{equation}\label{eqn:toy_linear_oed_optimization}
        \design\opt
          = \argmax_{\design \in [0, 1]^{\Nsens}} \,
                \logdet{ 
                    \sum_{i=1}^{3} {\mat{A}^{i}} \tran \ \wdesignmat(\design) \mat{A}^{i} 
                    + \Cparamprior\inv
                }
        \,,
    \end{equation}
    where the weighted precision matrix $\wdesignmat(\design)$ is given by~\eqref{eqn:pointwise_weighted_precision} 
    and the design weights are set to $w_j=\design_j,\, j=1, 2,\ldots, 5$. 
    The approach used to define these weights can be adjusted at any time by updating the OED configurations. 
    A method \code{update\_configurations()} is available through all DA and OED methods to enable updating most configurations through the life cycle of the object.
    Extensive validation is carried out upon instantiation and whenever the configurations are updated to make sure the settings of the object are consistent.
    
\begin{figure}[!htbp]
      \centering
        \includegraphics[width=0.450\textwidth]{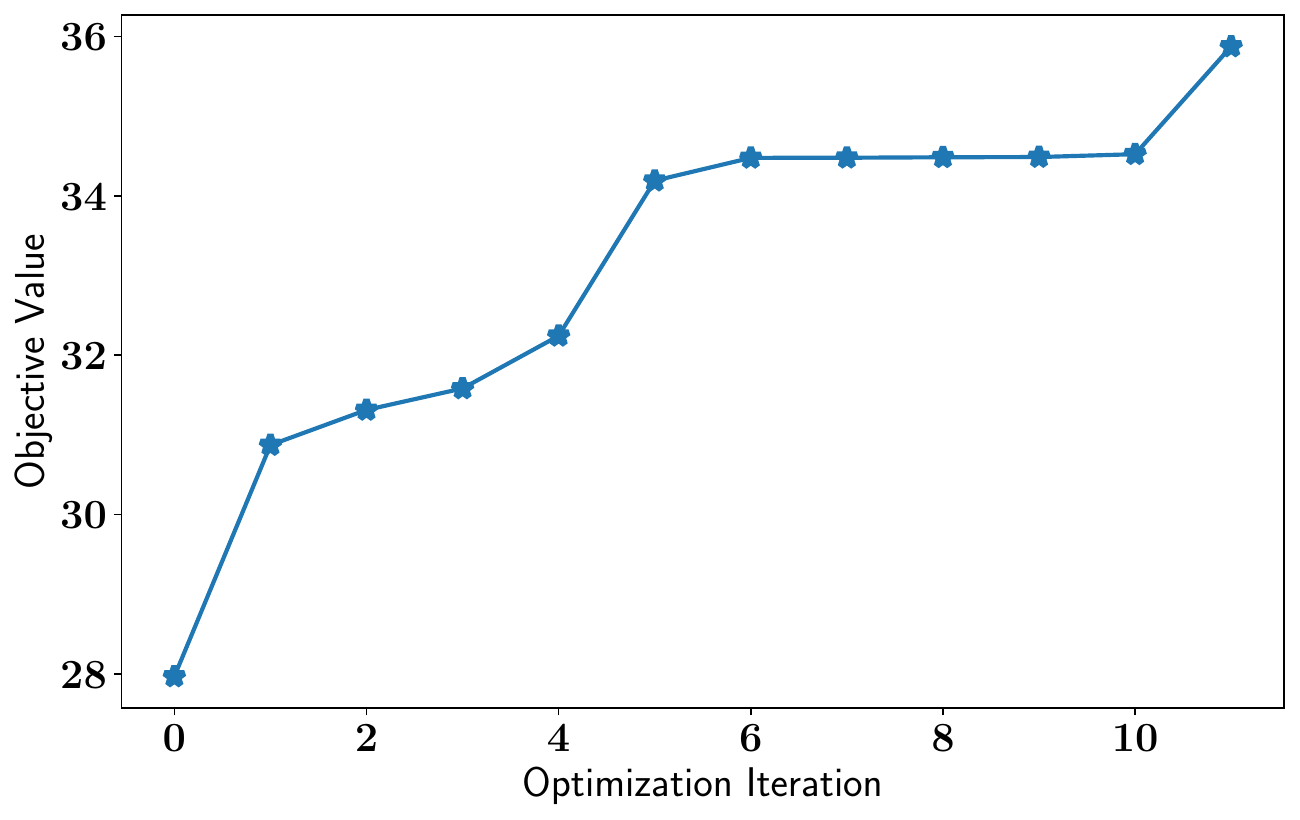}
        \includegraphics[width=0.460\textwidth]{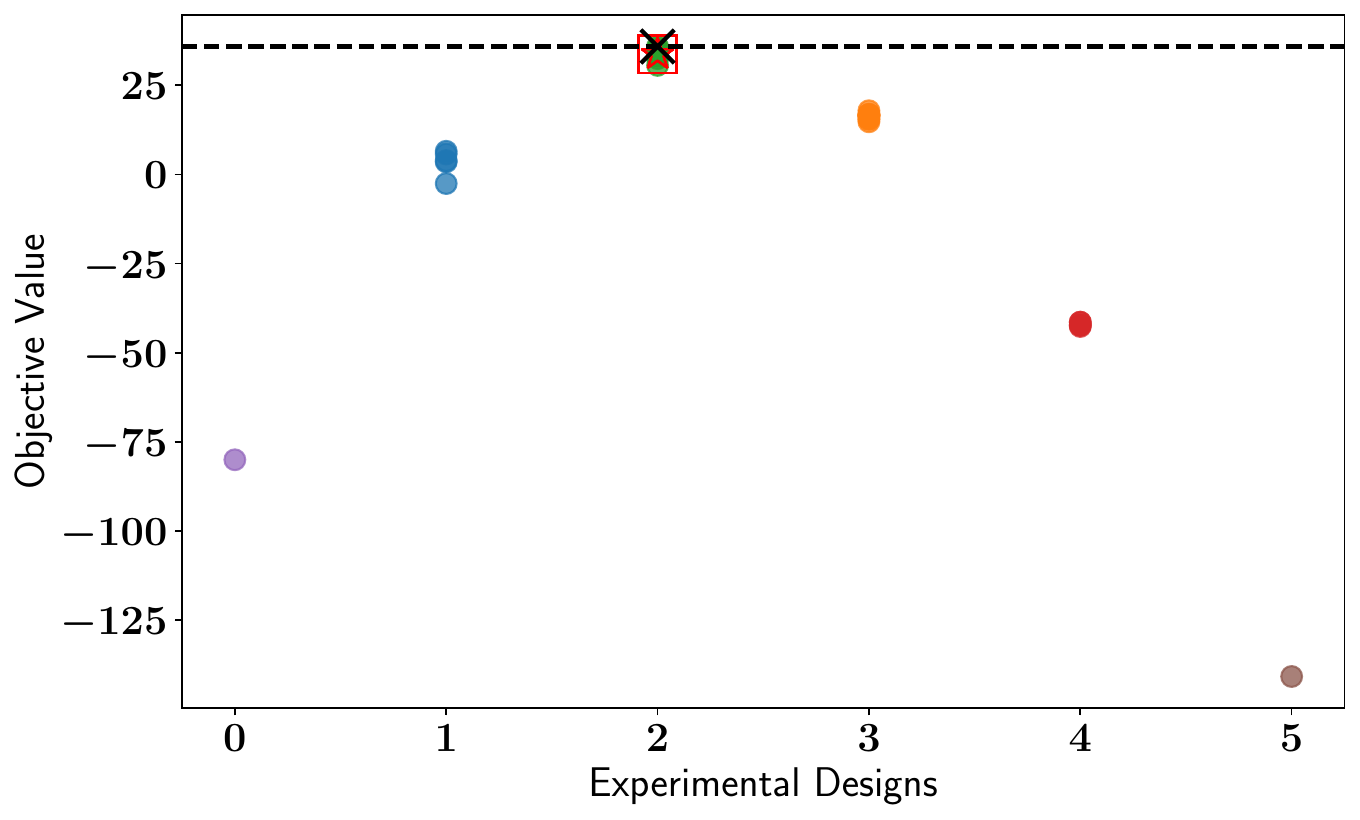}
        \includegraphics[width=0.6\textwidth]{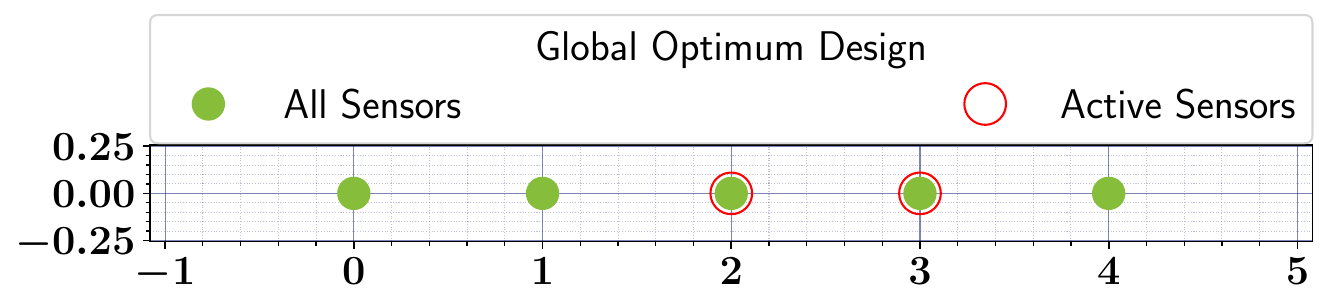}
      \caption{
        Subset of the plots generated by the \code{oed\_results.plot\_results} method after solving the OED problem described by \eqref{eqn:toy_linear_oed_optimization}.
        Top left: value of the utility (objective) function, i.e., the penalized OED criterion, over consecutive iterations of the optimization algorithm.
        Top right: value of the objective of the optimal solution (red star) returned by relaxed approach, compared with the global optimum solution (black $x$ mark), and all possible solutions marked as circles. 
        The x-axis denotes the indices of all possible binary designs from $1$ to $2^{\Nsens}=32$, and the y-axis displays the corresponding values of the optimization objective.
        Bottom: optimal design, showing optimal active sensors compared with all possible candidates (the 5 degrees of freedom).
      }
      \label{fig:ToyLinear_OED_Plots}
    \end{figure}

    The OED solver~\code{pyoed.oed.sensor\_placement.relaxed\_oed.SensorPlacementRelaxedOED} is used in this example
    to solve the OED optimization problem~\eqref{eqn:toy_linear_oed_optimization}. The solution process is invoked 
    by calling~\code{solve\_oed\_problem()}, a method associated with the created OED object,
    which returns an \code{OEDResults} object.
    The results object itself gives access to the OED problem and the underlying elements, 
    results of the OED problem, and built-in tools to visualize those results. 
    For example, by calling \code{plot\_results()} on the \code{OEDResults} object, several diagnostics plots are generated,
    including those displayed in Figure~\ref{fig:ToyLinear_OED_Plots}.
    These results show the performance of the optimization procedure as it proceeds over iterations, 
    compare the value of the optimal design with all possible designs (brute force),
    and display the optimal design over a grid of all possible candidate locations.
    The built-in visualization capabilities are simple, yet they are dynamic. 
    For example, the method~\code{plot\_results()} that generated the results in Figure~\ref{fig:ToyLinear_OED_Plots}
    adapts itself to the dimensionality of the problem and is capable of visualizing 1D, 2D, and 3D sensor configurations.

  \subsection{A standard model-constrained OED experiment}
  \label{subsec:results:standard_oed_examples}
    Now let us turn to the parameter identification for an advection-diffusion  model. 
    This is the foundation of an experiment widely used in the model-constrained OED literature for validating theoretical developments; see, for example,~\cite{alexanderian2016fast,AlexanderianPetraStadlerEtAl14,attia2018goal, VillaPetraGhattas18, attia2022optimal, attia2023robust}.
    Here we show how \pyoed can be used to solve and benchmark this OED problem, 
    thus providing a starting point for utilizing and developing multiple approaches for solving OED problems in general in \pyoed.
    The code employed for the discussion in this section 
    can be found in the \code{pyoed.examples.oed.OED\_AD\_FE} module with additional comments, details, and capabilities. 
    A corresponding Jupyter Notebook is also given in \code{pyoed.examples.tutorials}.
    
\begin{figure}[!htbp]
        \centering
          \includegraphics[width=0.25\textwidth]{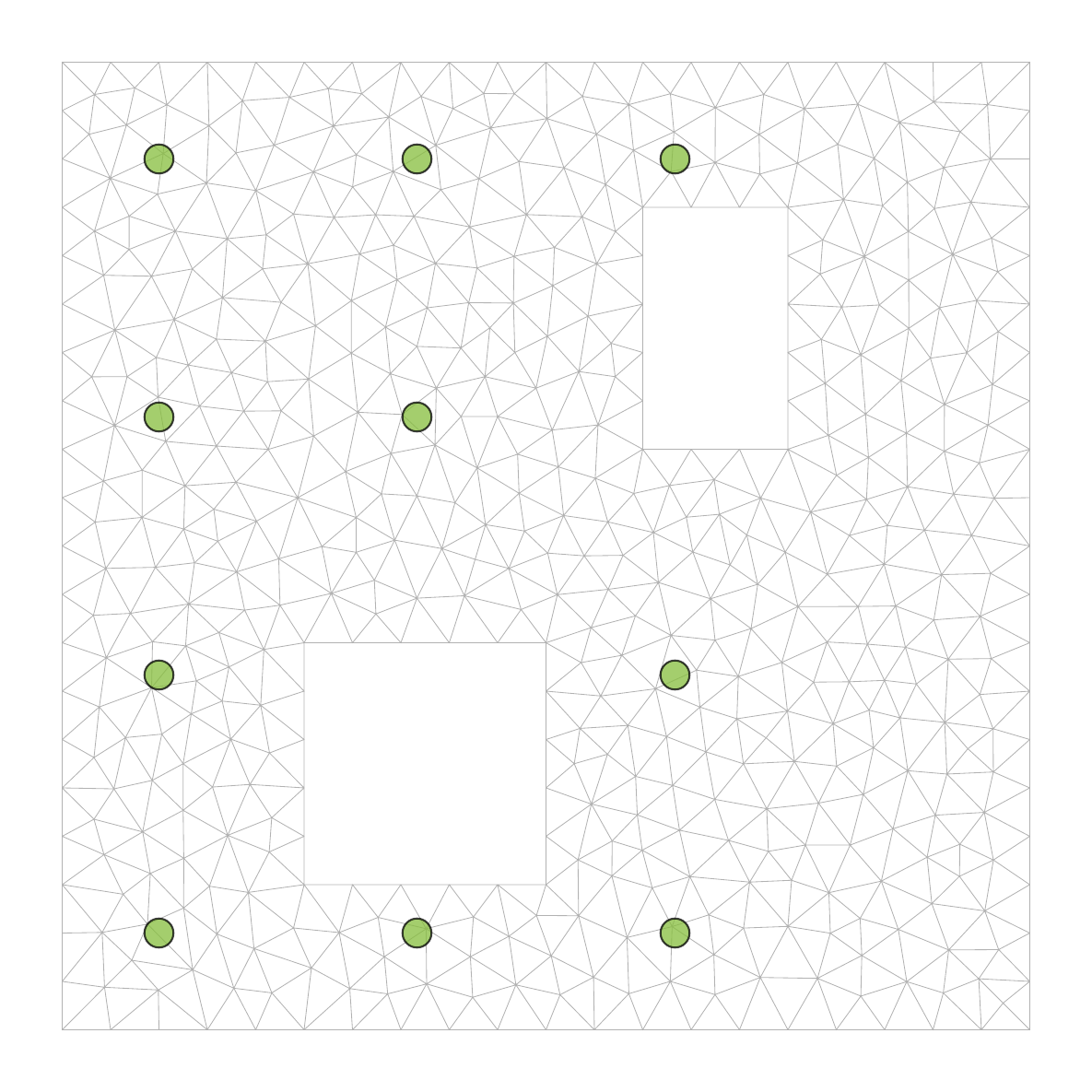}
          \qquad
          \includegraphics[width=0.25\textwidth]{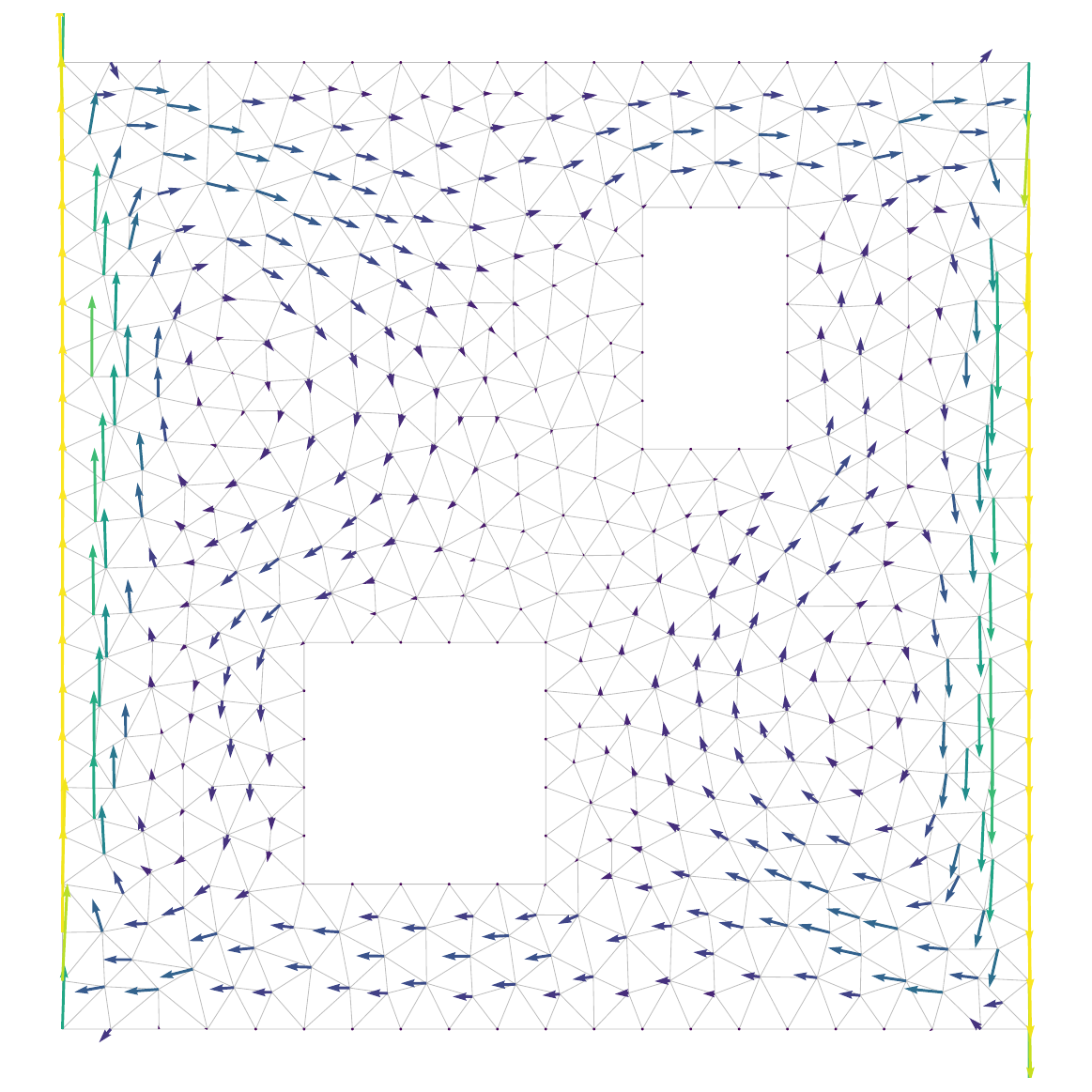}
          \caption{
            Left: finite elements discretization of the domain $\domain$ of the AD problem~\eqref{eqn:advection_diffusion}.
            Right: the velocity field $\vec{v}$.
          }
          \label{fig:AD_domain_velocity}
      \end{figure}

    \paragraph{The forward and inverse problems}
      The governing equations of the contaminant field $u = \xcont(\mathbf{x}, t)$ are given by 
      the following advection-diffusion equations and associated initial and boundary conditions:
      \begin{equation}\label{eqn:advection_diffusion}
        \begin{aligned}
          \xcont_t - \kappa \Delta \xcont + \vec{v} \cdot \nabla \xcont &= 0
            \quad \text{in } \domain \times [0,T],   \\
          \xcont(x,\,0) &= \theta \quad \text{in } \domain,  \\
          \kappa \nabla \xcont \cdot \vec{n} &= 0
            \quad \text{on } \partial \domain \times [0,T]\,,
        \end{aligned}
      \end{equation}
      where $\kappa>0$ is the diffusivity, $T$ is the simulation final time, and $\vec{v}$ is a predefined velocity field that is advecting the contaminant.
      The domain is the unit square $\domain:=(0, 1)^2$ with two rectangular regions modeling two buildings blocking flow.
      The velocity field $\vec{v}$ is obtained by solving a steady Navier--Stokes equation, with the side walls driving the flow, as detailed in~\cite{PetraStadler11,VillaPetraGhattas18,attia2022optimal}. 
      For the following numerical experiments, we use $\kappa=0.001$, and $T=5.5$. 
      The initial condition $\xcont(x,\,0)=\theta$ is assumed to be the unknown parameter we seek to infer. 
      \pyoed provides an implementation of this model in its core as the simulation model object
      \code{pyoed.models.simulation\_models.fenics\_models.AdvectionDiffusion2D}.
      The domain (with finite elements discretization) as well as the corresponding velocity field is shown in Figure~\ref{fig:AD_domain_velocity}.

      Now, turning to required error models, following~\cite{attia2018goal,PetraStadler11,VillaPetraGhattas18}, we impose a bi-Laplacian prior on the parameter $\GM{\iparb}{\Cparampriormat}$.
      Here $\Cparampriormat$ is a discretization of $\mathcal{A}^{-2}$, where $\mathcal{A}$ is a Laplacian-like operator.
      Regarding the observation configuration, a common choice for this model is to select some uniformly distributed candidate sensor locations; we will select $10$ such sensors. 
      Note that the process of constructing the operator that extracts a function in a finite element space onto the observation grid is nontrivial;
      hence, \pyoed provides a pointwise observation operator under \code{pyoed.models.observation\_operators.fenics\_observation\_operators)}.
      Additionally, we impose a Gaussian observational noise model with zero mean and diagonal covariance matrix with diagonal entries (variance) equal to $0.1$.
      Regarding the assimilation setup, as usual for a time-dependent simulation model, we construct a 4DVar instance similar to that employed in the first case presented in Section~\ref{subsec:ToyLinear}.
      Synthetic observations (data) are obtained by integrating the simulation model using a synthetic true initial condition, followed by the application of the observation operator and contaminating the results with additive Gaussian noise.
      One can solve the inverse problem to infer the inference parameter exactly as in Section~\ref{subsec:ToyLinear}; however, 
      we focus here on solving a sensor placement OED problem. 
  
    \paragraph{The OED problem}
      While we have provided a bespoke introduction for the contaminant source identification assimilation procedure,
      the following OED process is general to any model-constrained assimilation process.
      For illustration, we seek to activate only $4$ sensors out of the candidate $10$, 
      and we use the trace of the Fisher information matrix ($\FIM$) as the utility function. 
      To enforce the budget, we use an $\ell_0$ penalty term on the OED optimization objective, and we employ the stochastic OED approach 
      detailed in~\cite{attia2022stochastic}.
      The stochastic OED optimization objective~\eqref{eqn:stochastic_OED_optimization} 
      in this case takes the form
      \begin{equation}\label{eqn:stochastic_OED_optimization_AD}
        \hyperparam\opt
          = \argmax_{\hyperparam\in[0,1]^{\Nsens}}
              \Expect{\design\sim\CondProb{\design}{\hyperparam}}{\trace[\FIM(\design)]- \baseline - \regpenalty \Big|\wnorm{\design}{0} - 4 \Bigr| } \,, 
      \end{equation}
      where the regularization parameter $\regpenalty>0$ is set by trial and error, here $1e+4$, and the baseline $\baseline$ is calculated automatically by the stochastic OED 
      object. Specifically, \code{pyoed.oed.sensor\_placement.binary\_oed.SensorPlacementBinaryOED} implements the stochastic approach~\cite{attia2022stochastic} and is used here.
      
\begin{figure}[!htbp]
       \centering
         \includegraphics[width=0.35\textwidth]{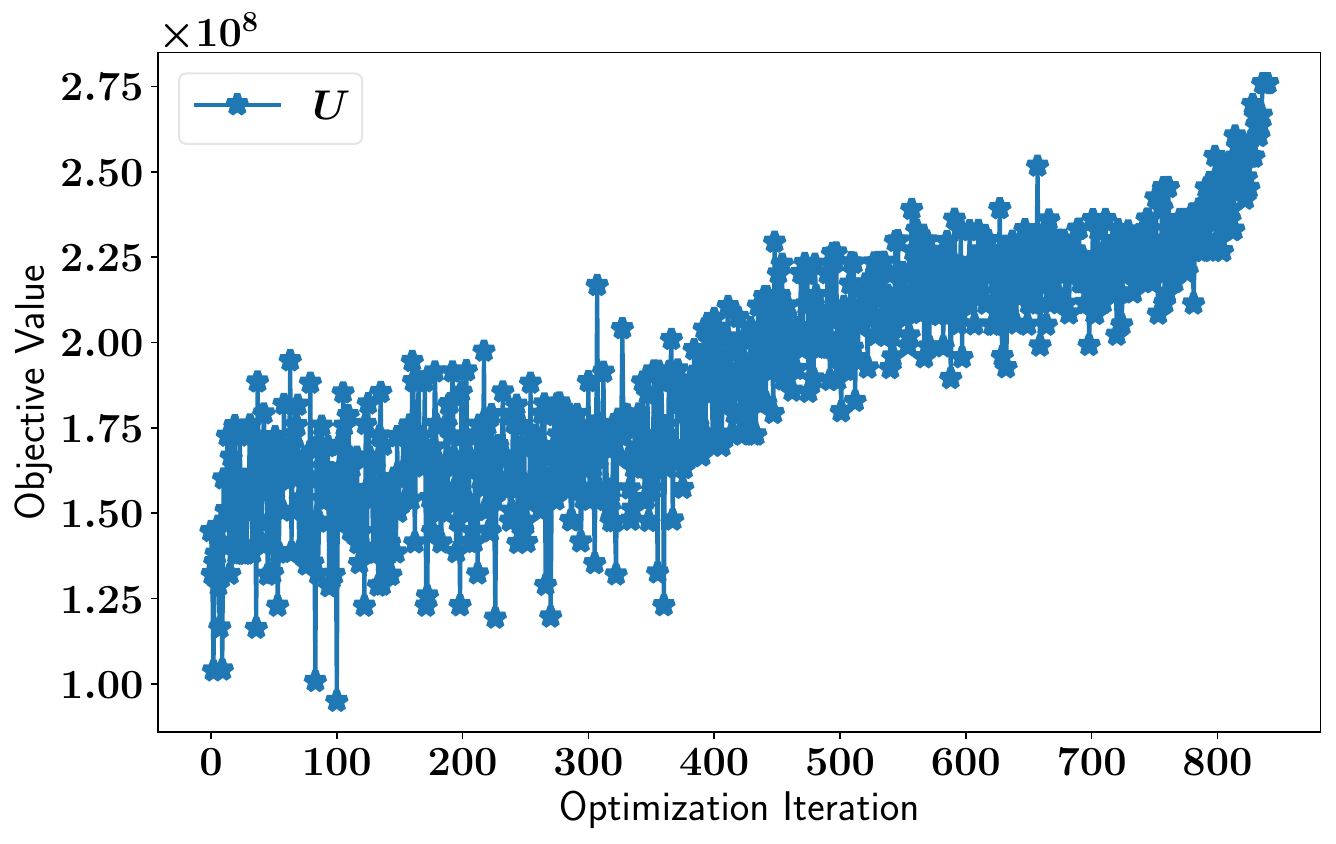}
         \quad 
         \includegraphics[width=0.22\textwidth]{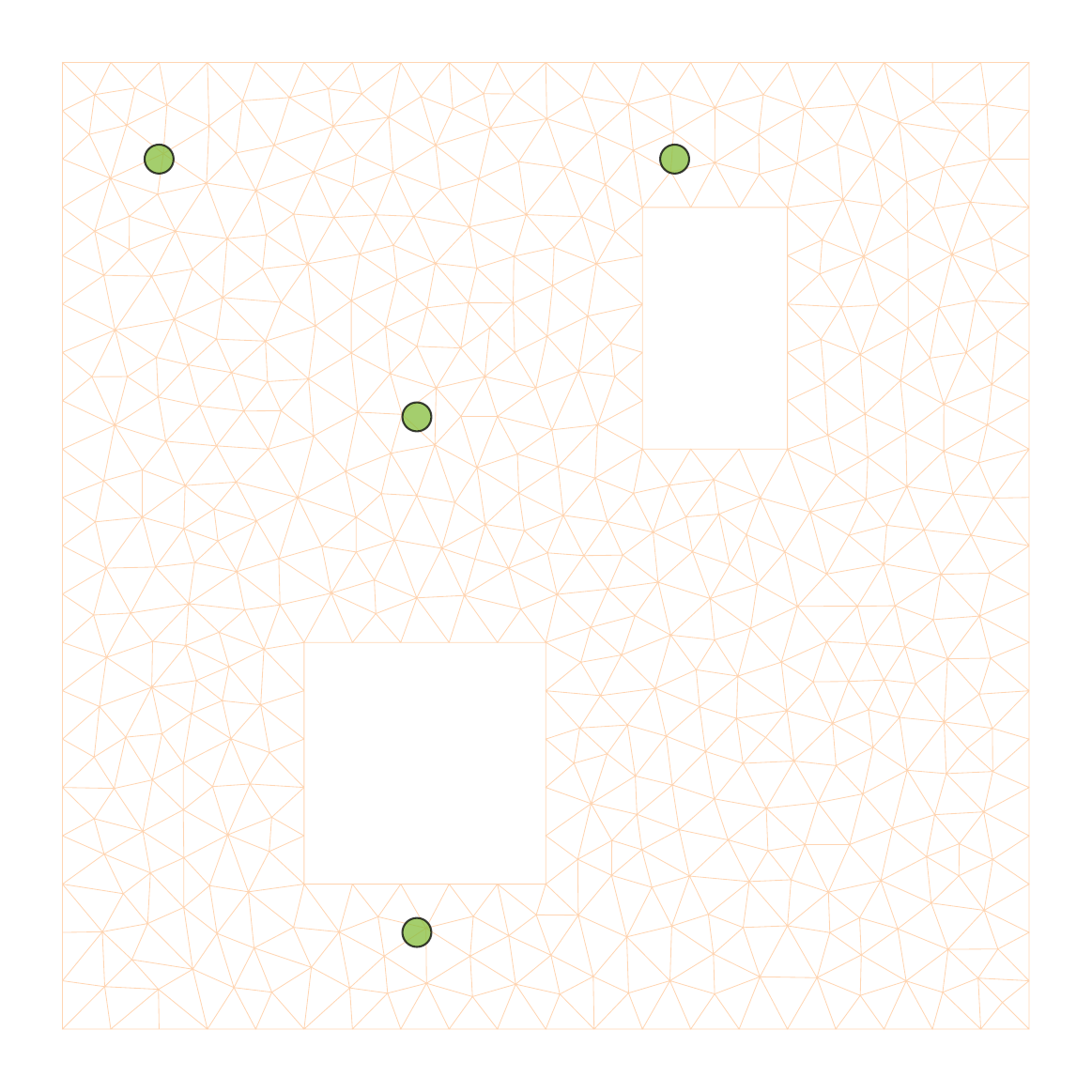}
         \quad
         \includegraphics[width=0.35\textwidth]{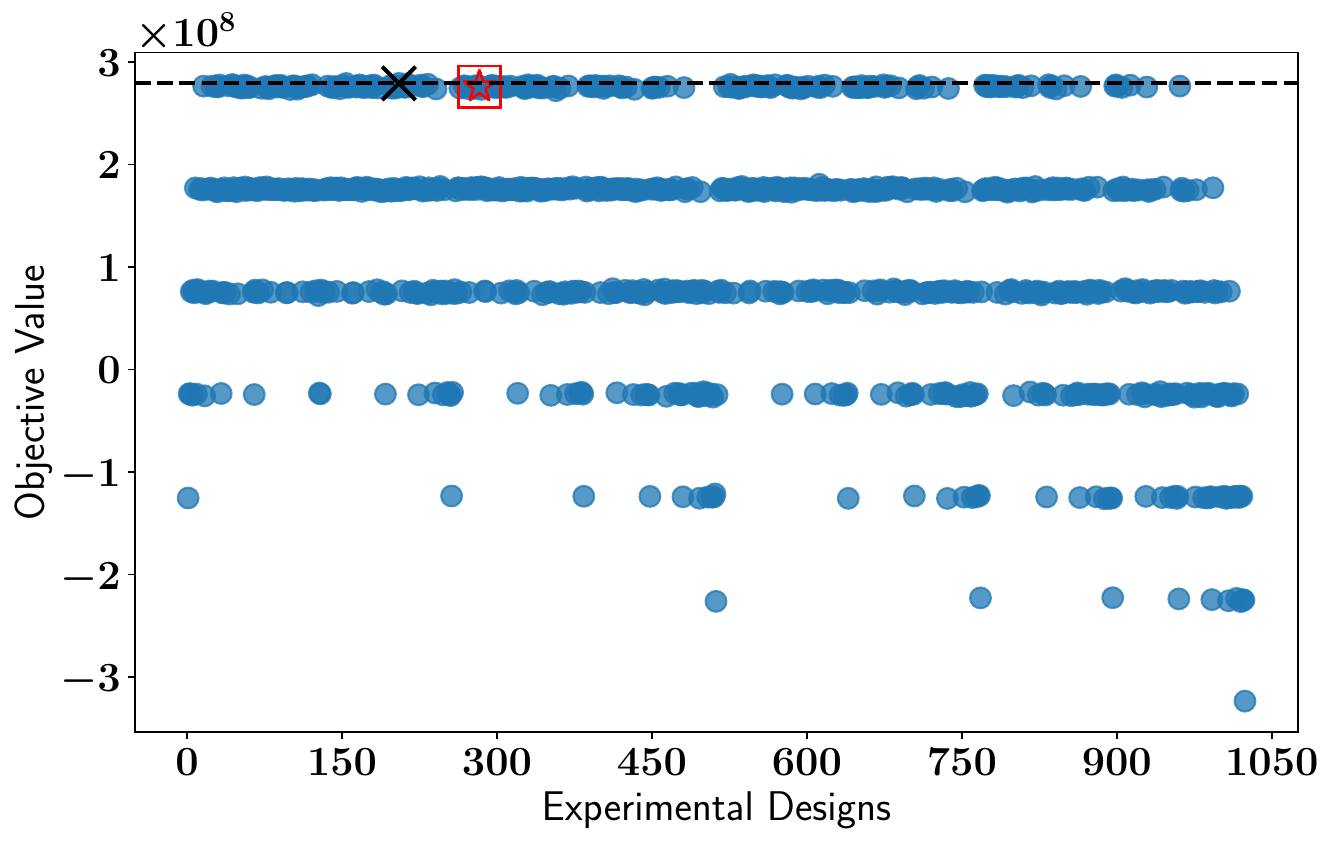}
       \caption{Subset of the plots generated by the \code{oed\_problem.plot\_results} method.
         Left: value of the utility (objective) function, i.e., the penalized OED criterion, over consecutive iterations of the optimization algorithm.
         Middle: optimal solution, showing optimal sensor locations in the domain.
         Right: value of the objective of the optimal solution (red star) returned by the stochastic learning approach~\cite{attia2022stochastic}, 
         compared with the global optimum solution (black $x$ mark) 
         and all possible solutions (marked as blue circles); the x-axis shows the indexes of all possible binary designs from $1$ to $2^{\Nsens=10}=1024$, 
         and the y-axis shows the corresponding values of the optimization objective.
       }
       \label{fig:AD_FE_Plots}
     \end{figure}

      We can then solve the OED problem and compare the results against the brute force enumeration of all possible designs; see Figure~\ref{fig:AD_FE_Plots} for the results.
      Visualization of the OED results and the capability for generating brute force results are  provided through the \code{OEDResults} generated by solving the OED problem.
      A partial set of results of this experiment is shown in Figure~\ref{fig:AD_FE_Plots}.
      We note that both the quality of observation configuration and the budget constraint improve over iterations. 
      The brute force search, which scans all $2^{10}=1024$ possible designs/configurations, 
      verifies that the \pyoed result using the stochastic learning approach is close to the global optimum solution while consuming significantly fewer resources.
      This analysis is not necessarily optimal, and \pyoed aims to help researchers benchmark, analyze, and improve such approaches.
     %

\section{Concluding Remarks}
\label{sec:Conclusions}
  This work describes \pyoed, a highly extensible high-level software package for OED in inverse problems and DA.
  \pyoed aims to be a comprehensive Python toolkit for model-constrained OED.
  The package targets scientists and researchers interested in understanding the details of OED formulations and approaches. 
  It is also meant to enable researchers to experiment with standard and innovative OED technologies within external test problems (e.g., simulations).
  As the community continues to progress in  understanding  Bayesian inversion, DA, and OED, we intend to extend \pyoed alongside them.
  As an example, we are paying close  attention to advances in automatic differentiation \cite{james2018jax} and numerical methods 
  for the accelerated evaluation of previously intractable utility criteria \cite{beck2018fast}.
  While we focused the discussions in this paper on specific OED approaches, the current version \pyoed provides several other implementations and emphasizes implementing the essential infrastructure that enables combining DA and OED elements with other parts of the package.
  Thus, we plan to expand the package with additional DA methods such as variants of the Kalman filtering and smoothing algorithms, 
  particle filtering, and hybrid variational-ensemble methods.
  Relevant optimization methods such as branch and bound will also be added.
  The main limitation of the initial version of \pyoed is scalability.
  Specifically, the concept is developed without parallelization capability.
  In future versions of \pyoed, scalability will be achieved by adding Message Passing Interface (MPI) support, for example using the \code{mpi4py} package, and by supporting \code{PETSc}~\cite{balay2022petsc}.
  Performance will also be enhanced by converting or rewriting suitable parts of the package in \code{Cython}.



\begin{acks}
    This material is based upon work supported by the U.S. Department of Energy, Office of Science, under contract number DE-AC02-06CH11357.
    This work was supported in part by the U.S. Department of Energy, Office of Science, Office of Advanced Scientific Computing Research and Office of Nuclear Physics, Scientific Discovery through Advanced Computing (SciDAC) Program through the FASTMath Institute.
    The work of SEA was supported by Argonne National Laboratory during his appointment as a 2021 Wallace Givens Associate.
\end{acks}

\bibliographystyle{ACM-Reference-Format}
\bibliography{references}

\iftrue
\null \vfill
  \begin{flushright}
  \scriptsize \framebox{\parbox{3.2in}{
  The submitted manuscript has been created by UChicago Argonne, LLC,
  Operator of Argonne National Laboratory (``Argonne"). Argonne, a
  U.S. Department of Energy Office of Science laboratory, is operated
  under Contract No. DE-AC02-06CH11357. The U.S. Government retains for
  itself, and others acting on its behalf, a paid-up nonexclusive,
  irrevocable worldwide license in said article to reproduce, prepare
  derivative works, distribute copies to the public, and perform
  publicly and display publicly, by or on behalf of the Government.
  The Department of
  Energy will provide public access to these results of federally sponsored research in accordance
  with the DOE Public Access Plan. http://energy.gov/downloads/doe-public-access-plan. }}
  \normalsize
  \end{flushright}
\fi

\end{document}